\documentclass[journal,11pt,draftclsnofoot,onecolumn]{IEEEtran}
\usepackage{amsmath,amsfonts}
\usepackage{algorithmic}
\usepackage{algorithm}
\usepackage{array}
\usepackage[caption=false,font=normalsize,labelfont=sf,textfont=sf]{subfig}
\usepackage{textcomp}
\usepackage{stfloats}
\usepackage{url}
\usepackage{verbatim}
\usepackage{graphicx}
\usepackage{cite}
\usepackage{mathptmx}
\usepackage{etoolbox}
\usepackage{upgreek}

\begin{document}

\title{Ultrafast and precise distance measurement via real-time chirped pulse interferometry}

\author{Mingyang Xu, Hanzhong Wu, Jiawen Zhi, Yang Liu, Jie Zhang, Zehuang Lu, and Chenggang Shao
\thanks{This work is supported by National Key Research and Development Program of China (2022YFC2204601); National Natural Science Foundation of China (11925503, 12275093); Natural Science Foundation of Hubei Province (2021CFB019), and State Key Laboratory of applied optics (SKLAO2022001A10). (Corresponding authors: Hanzhong Wu; Zehuang Lu, and Chenggang Shao.)}
\thanks{Mingyang Xu, Jiawen Zhi, Jie Zhang, Zehuang Lu, and Chenggang Shao are with the MOE Key Laboratory of Fundamental Physical Quantities Measurements, Hubei Key Laboratory of Gravitation and Quantum Physics, PGMF and School of Physics, Huazhong University of Science and Technology, Wuhan 430074, China  (e-mail:
mingyangxu@hust.edu.cn; d202280063@hust.edu.cn; jie.zhang@mail.hust.edu.cn; zehuanglu@hust.edu.cn; cgshao@hust.edu.cn). }
\thanks{Hanzhong Wu is with the MOE Key Laboratory of Fundamental Physical Quantities Measurements, Hubei Key Laboratory of Gravitation and Quantum Physics, PGMF and School of Physics, Huazhong University of Science and Technology, Wuhan 430074, China, and is also with the State Key Laboratory of Applied Optics, Changchun Institute of Optics, Fine Mechanics and Physics, Chinese Academy of Sciences, Changchun 130033, China  (e-mail:
wuhanzhong@hust.edu.cn). }
\thanks{Yang Liu is with the National Institute of Metrology, Beijing 100013, China (e-mail:
liuyang1@nim.ac.cn). }

}



\maketitle

\begin{abstract}
Laser frequency combs, which are composed of a series of equally-spaced coherent frequency components, have triggered revolutionary progress for precision spectroscopy and optical metrology. Length/distance is of fundamental importance in both science and technology. In this work, we describe a ranging scheme based on chirped pulse interferometry. In contrast to the traditional spectral interferometry, the local oscillator is strongly chirped which is able to meet the measurement pulses at arbitrary distances, and therefore the dead zones can be removed. The distances can be precisely determined via two measurement steps based on time-of-flight method and synthetic wavelength interferometry, respectively. To overcome the speed limitation of the optical spectrum analyzer, the spectrograms are stretched and detected by a fast photodetector and oscilloscope, and consequently mapped into the time domain in real time. The experimental results indicate that the measurement uncertainty can be well within ±2 $\upmu$m, compared with the reference distance meter. The Allan deviation can reach 0.4 $\upmu$m at averaging time of 4 ns, 25 nm at 1 $\upmu$s, and can achieve 2 nm at 100 $\upmu$s averaging time. We also measure a spinning disk with grooves of different depths to verify the measurement speed, and the results show that the grooves with about 150 m/s line speed can be clearly captured. Our method provides a unique combination of non-dead zones, ultrafast measurement speed, high precision and accuracy, large ambiguity range, and with only one single comb source. This system could offer a powerful solution for the field measurements in practical applications in future.
\end{abstract}

\begin{IEEEkeywords}
absolute distance measurement; frequency comb; chirped pulse interferometry.
\end{IEEEkeywords}

\section{introduction}
\noindent During the past two decades, laser frequency combs have achieved huge success in a vast number of applications, e.g., absolute frequency measurement, precision spectroscopy, optical clock, communications, astronomy, and absolute distance measurement, etc {\cite{1,2}}. This kind of comb laser features a series of coherent peaks with equal space in the frequency domain, and exhibits a pulsed nature in the time domain. There are two degrees of freedom, the repetition frequency $f_{rep}$ and the carrier-envelope-offset frequency $f_{ceo}$, to characterize the laser frequency combs, and one single line can be expressed as $m$×$f_{rep}+f_{ceo}$. $m$ is an integer. As long as $f_{rep}$ and $f_{ceo}$ are locked to the external frequency reference, laser frequency combs can serve as a precise and accurate ruler both in the frequency and time domains {\cite{3}}. Several mechanisms are able to stimulate frequency combs, such as passive mode-locked lasers, electro-optic modulation {\cite{4,5}}, and nonlinearity evolution in micro resonators {\cite{6,7}}. To date, all these kinds of laser frequency combs can be self referenced to an external frequency reference.

Since the first demonstration in 2000 {\cite{8}}, frequency comb-based absolute distance measurement have developed for over 20 years, and various schemes have been investigated in great depth, such as inter-mode beat {\cite{8,9,10}}, pulse cross correlation {\cite{11,12,13,14,15,16,17,18}}, dual comb interferometry {\cite{19,20,21,22,23,24,25,26,27,119}}, multi-heterodyne interferometry {\cite{28}},  pulse-to-pulse alignment {\cite{29,30}}, and stationary phase evaluation {\cite{31,32}}. Additionally, frequency comb can also work as the reference source when using one or more continuous wave lasers to measure the absolute distances. The problem may be that the whole system involves multi optical sources to increase the system cost and complexity {\cite{33,34,35}}. Very recently, optical ranging with microcombs has attracted the interest of the scientists due to their small footprint and low power consumption. All the methods mentioned above can be used to measure the distances using the soliton microcombs, and the attempts of the pulse cross correlation {\cite{18}}, dual-comb interferometry {\cite{36,37}}, and dispersive interferometry have already been reported {\cite{38,39}}. To date, the physics of the microcombs is becoming clear, but the long-term reliable operation of the soliton state is still relatively challenging. In addition, the commercialization of the microcombs needs to be accelerated. For any kind of measurement methods, there are several parameters to evaluate their performances, including precision/accuracy, ambiguity range, measurement speed, and system cost {\cite{20,30}}. Generally speaking, improvement in one parameter is achieved at the exchange of degradation in other parameters.

\begin{figure*}[htbp]
\centering
    \includegraphics[width=0.75\textwidth]{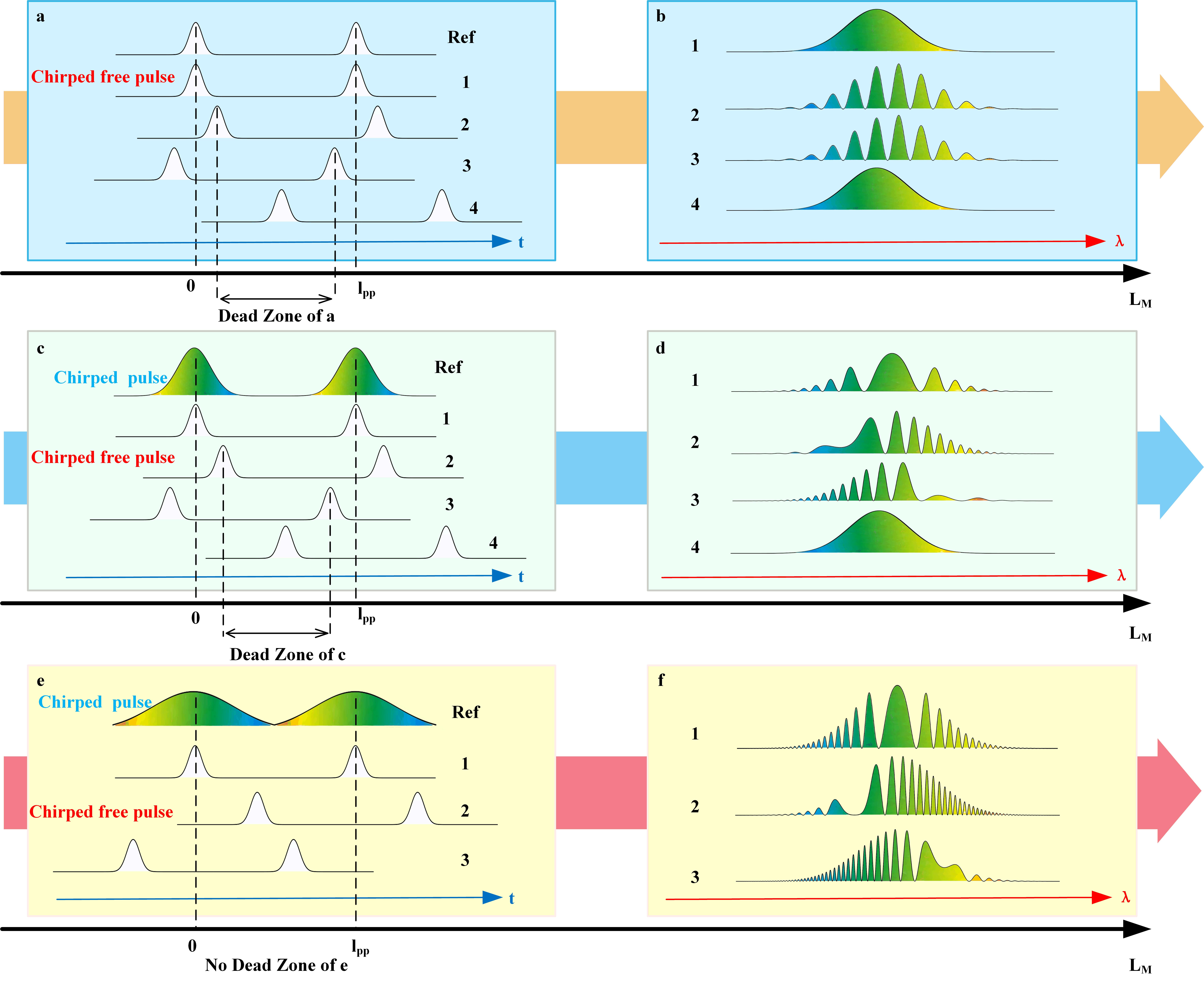}
    \caption{ Frequency comb interferometry in the frequency domain. (a) and (b): In the traditional spectral interferometry, the spectrum will be modulated, and the modulation frequency is constant along the wavelength axis, e.g., the situations of 2 and 3. Please note that, the spectrograms corresponding to the situations of 2 and 3 are the same, despite that the distances are different. Limited by the resolution of the optical spectrum analyzer, the spectrograms {can not} be reconstructed when the distance is too large, e.g., the situation of 4; (c) and (d): Chirped pulse interferometry will occur when the reference pulses are chirped. The modulation frequency of the spectrograms is not constant, e.g., the situation of 1, and there is a widest fringe. When the distances are changed, the position of the widest fringe changes, e.g., the situations of 2 and 3. Note that, the spectrograms corresponding to the situations of 2 and 3 are not the same. Similarly, the spectrograms {can not} be reconstructed when the distance is too large, e.g., the situation of 4; (e) and (f): The reference pulse is strongly chirped, and the pulse width can well cover the pulse-to-pulse interval. The modulation frequency of the spectrograms is larger, and there is no dead zone along the measurement path.}
    \label{fig:1}
\end{figure*}

One important group among these methods is the spectral interferometry, also known as dispersive interferometry, in which the spectrum will be modulated when the pulses interferes with each other {\cite{40}}. In this case and often, an optical spectrum analyzer (OSA) is exploited to measure and hold the spectrograms. The following data process including the Fourier transform and the phase unwrapping can extract the distance information. Yet, limited by the resolution of the optical spectrum analyzer, the spectrograms can not be reconstructed any more when the pulse-to-pulse interval is too large {\cite{100,101}}.  This means that, there are dead zones along the measurement path, as shown in {Figure \ref{fig:1}}(a). To overcome this limitation, virtual imaging phase array VIPA and a grating can be used to disperse the light in a two-dimensional plane, and a CCD camera can consequently image each individual mode {\cite{41}}. This scheme can be considered as homodyne interferometry with a number of wavelengths, and the dead zones can be therefore removed. But, frequency comb with larger repetition frequency (e.g., 1 GHz) is needed, otherwise the individual mode can not be resolved. It is necessary to mention that the problem of the dead zone can be relaxed if using the electro-optic combs {\cite{4,25}} or the micro combs {\cite{36,37}},  since the repetition frequencies of these combs are inherently large over tens of GHz. Therefore, the pulse-to-pulse interval is relatively small. Additionally, for spectral interferometry, the small distances of +$l$ and -$l$ in the vicinity of $N$×$L_{pp}$ can not be directly distinguished {\cite{40}}.  + and – represent the positions before and after the reference pulse, $N$ is an integer, and $L_{pp}$ is the pulse-to-pulse length of the frequency comb.

Another problem of using the CCD-based OSA is that the measurement speed is limited by the frame rate of the camera, which is only about kfps in general. Similarly, the speed of the OSA based on the long translation stage is determined by the moving speed of the mechanical stage {\cite{31}}. In recent years, a new kind of real-time OSA has found a number of applications, which relies on the theory of dispersive Fourier transform {DFT} {\cite{42,43}}, also referred to as time stretch. In DFT, the femtosecond pulses are greatly stretched, and then detected by a fast photodetector and a fast AD card. Now, technique of time stretch has been used in the fields of {optical ranging} {\cite{44,102}}, {imaging} {\cite{45}}, {spectroscopy} {\cite{103,104,105}}, {intelligent control of the pulses} {\cite{106}}, {and ultrafast phenomenon observation} {{\cite{46,107,108,109,110,111}}}. In the case of long distance measurement using time stretch-based dispersive interferometry, the spectrograms in the frequency domain can be translated into the time domain, and more important the measurement speed is directly determined by the repetition frequency of the pulsed laser, which can be as high as several nanoseconds per measurement {\cite{112,113}}. However, limited by the bandwidth of the photodetector and the AD card, the spectrograms can not be reconstructed either for this new type of OSA when the distance is too large. This means that, the dead zones still exist in the measurement path, as shown in {{Figure \ref{fig:1}}(a)}.

\begin{figure}[htbp]
\centering
    \includegraphics[width=0.5\textwidth]{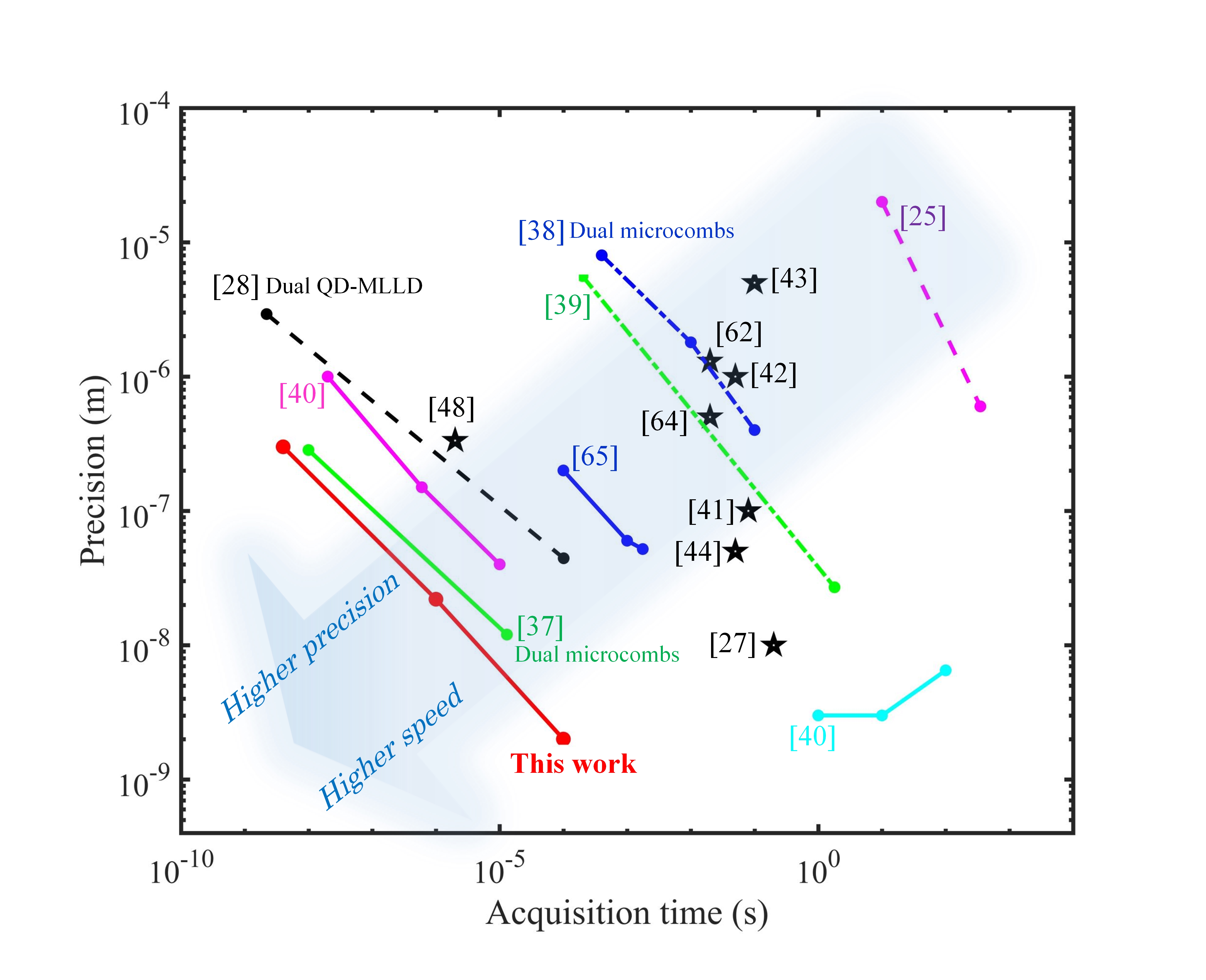}
    \caption{Recent progress in the field of distance measurement. Please note that, here we focus on the methods of spectral interferometry and chirped pulse interferometry. Some nice reviews can be seen in \cite{116,117,118}.}
    \label{fig:n2}
\end{figure}

If the reference pulses are chirped by a dispersive element (e.g., a piece of fiber, a pair of gratings, or a piece of glass), chirped pulse interferometry will occur {\cite{47,48}}. Different from the classical spectral interferometry, the modulation frequency of the spectrograms for chirped pulse interferometry is not constant any more {\cite{114,115}}. There is a widest fringe in the spectrograms, where the arm length difference is balanced at the corresponding wavelength {\cite{49}}. Compared to the spectral interferometry, the measurable range is expanded. However, the dead zones still exist due to the relatively small dispersion introduced in the reference arm, as shown in {Figure \ref{fig:1}(c)}. If the dispersion amount is sufficiently large, the measurement pulse can always meet the reference pulse in space, and the dead zones can be thus removed in principle, as shown in {Figure \ref{fig:1}(e)}. Considering the measurement speed, we can also use the technique of DFT in chirped pulse interferometry. Nevertheless, the limitation of the bandwidth of the photodetector and the AD card is not very strict, since a widest fringe exists in the spectrograms, whose width keeps constant for different distances. In contrast, for the spectral interferometry, the modulation frequency linearly increases with increasing the distance. Here, we summarize the recent progress for the field of distance measurement in Figure \ref{fig:n2}, mostly focusing on the spectral interferometry and chirped pulse interferometry. Some reviews can be found in \cite{116,117,118}.

In this work, we describe a method based on real-time chirped pulse interferometry for absolute distance measurement. A highly dispersive fiber is used to broaden the reference pulse. To make sure that the measurement pulses can always overlap with the reference pulse, we change the repetition frequency to adjust the relative delay between the measurement and the reference pulses due to the difference of the pulse index. Therefore, the spectrograms can be always observed along the whole measurement path, and the dead zones can be removed. The distances can be determined via two steps of phase measurements, which are successively the flight time of the pulses and the phase of the synthetic wavelength. In the detection of the spectrograms, we use the highly dispersive fiber to stretch the pulse, and real-time detect the spectrograms with the help of fast photodetectors and oscilloscope. In this case, the measurement speed is directly linked to the repetition frequency, up to about 250 MHz (4 ns for one single measurement). The experimental results show that the distances can be measured with high accuracy and precision, high measurement speed, and high resolution. This paper is organized as follows: Section 2 introduces the measurement principle. The distances can be measured via the slope of the phase difference between the two spectrograms, even if the phase of one spectrogram is not linear any more. The experimental setup is described in Section 3. Section 4 demonstrates the experimental results, in the long distance measurement and ultrafast measurement, respectively. Uncertainty budget is discussed in Section 5. Finally, we summarize this work in Section 6.

\section{Measurement principle}
\begin{figure}[htbp]
\centering
    \includegraphics[width=0.5\textwidth]{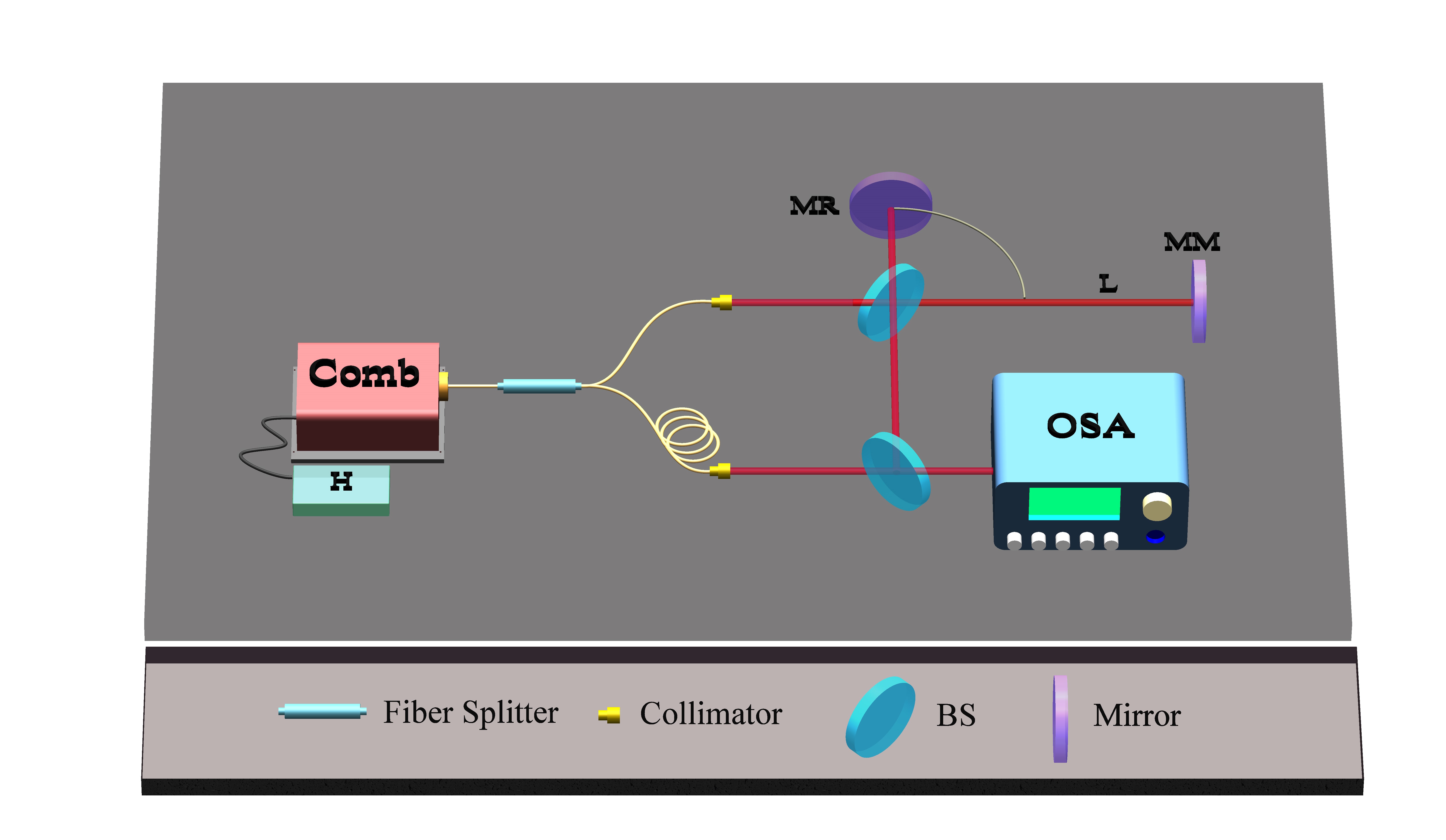}
    \caption{ Schematic of the measurement principle of the chirped pulse interferometry. H: Hydrogen maser; MR: reference mirror; MM: measurement mirror; BS: beam splitter; OSA: optical spectrum analyzer. The local oscillator is greatly broadened by a highly dispersive fiber.}
    \label{fig:2}
\end{figure}

\noindent
The schematic of the measurement principle is shown in {Figure \ref{fig:2}}. Laser frequency comb, which is locked to the Hydrogen maser, is split into two parts. One part is strongly broadened by a highly dispersive fiber, named as local oscillator, and the other part injects into a Michelson interferometer. The pulses reflected by the reference mirror MR are named as reference pulse, and that reflected by the measurement mirror are measurement pulses. The distance measured is the optical path difference between the measurement arm and the reference arm. The pulses are combined at a beam splitter, and finally detected by an optical spectrum analyzer. The obtained spectrograms can be used to process to measure the distances.

Assume that, the spectrum of the frequency comb is $E(\omega)$. The reference pulse $E_{ref}(\omega)$ can be expressed as $\alpha$$E(\omega)$ in the frequency domain, and $\alpha$ is the power factor. The measurement pulse $E_{meas}(\omega)$ can be written as $\beta$$E(\omega)exp(-i\tau\omega)$, and $\beta$ is the power factor corresponding to the measurement pulse. $\tau$ is the time delay between the reference pulse and measurement pulse.

The local oscillator $E_{LO}(\omega)$ can be expressed as $\gamma$$E(\omega)\cdot{exp(i\psi)}$, where $\gamma$ is the power factor, and $\psi$ is the phase change caused by the long dispersive fiber, which can be Taylor expanded as:
\begin{equation} \label{eq:1}
\begin{split}
\psi(\omega)&=\sum_{m=1}^{\infty} \frac{\beta_m}{m !}\left(\omega-\omega_c\right)^m z\\
&=\beta_1\left(\omega-\omega_c\right)\!z\!+\!\frac{\beta_2}{2}\left(\omega-\omega_c\right)^2z\!+\!\frac{\beta_3}{6}\left(\omega-\omega_c\right)^3z \ldots
\end{split}
\end{equation}
where $\beta_m$ is the coefficient of the group velocity dispersion at different orders, $m$ is an integer, $\omega_c$ is the center angular frequency, and $z$ is the fiber length. The spectrogram corresponding to the reference pulse can be expressed as:
\begin{equation} \label{eq:2}
\begin{split}
I_{ref}\left( \omega \right) =&\left( E_{ref}\left( \omega \right) +E_{LO}\left( \omega \right) \right) ^2\\
=&E_{ref}\left( \omega \right) ^2\!+\!E_{LO}\left( \omega \right) ^2\!+\!2Re\!\:\left[ E_{ref}\left( \omega \right) E_{LO}^{*}\left( \omega \right) \right] \\
=&E_{ref}\left( \omega \right) ^2+E_{LO}\left( \omega \right) ^2\\
&+2\left| E_{ref}\left( \omega \right) \right|\left| E_{LO}\left( \omega \right) \right|cos\!\:\left( \phi _r-\psi \right) \\
=&E\left( \omega \right) ^2\left[ \alpha ^2+\gamma ^2+2\alpha \gamma cos\left( \phi _r-\psi \right) \right] 
\end{split}
\end{equation}
where $E_{ref}(\omega)$ is the spectrum of the reference pulse, and $\phi_r$ is the initial phase difference between the reference pulse and the local oscillator. Based on Equations (\ref{eq:1}) and (\ref{eq:2}), we find that, different from the classical spectral interferometry, the spectral phase in the chirped pulse interferometry is not linearly related to $\omega$, which means that the modulation frequency of the spectrograms is not constant. There will be a “widest fringe” in the spectrograms. Similarly, the spectrogram corresponding to the measurement pulse can be written as:
\begin{equation} \label{eq:3}
\begin{split}
I_{meas}\left( \omega \right) =&E_{meas}\left( \omega \right) ^2+E_{LO}\left( \omega \right) ^2\\
&+2\left| E_{meas}\left( \omega \right) \right|\left| E_{LO}\left( \omega \right) \right|cos\!\:\left( \phi _r-\tau \omega -\psi \right) \\
=&E\left( \omega \right) ^2\left[ \beta ^2+\gamma ^2+2\beta \gamma cos\!\:\left( \phi _r-\tau \omega -\psi \right) \right] 
\end{split}
\end{equation}
where $E_{meas}(\omega)$ is the spectrum of the measurement pulse. Similar with the principle of dispersive interferometry in Ref. \cite{40}, the phase change from Equation (\ref{eq:2}) to Equation (\ref{eq:3}) can be given by:
\begin{equation} \label{eq:4}
\Delta \phi =\tau \omega 
\end{equation}

\begin{figure*}[htbp]
\centering
    \includegraphics[width=0.8\textwidth]{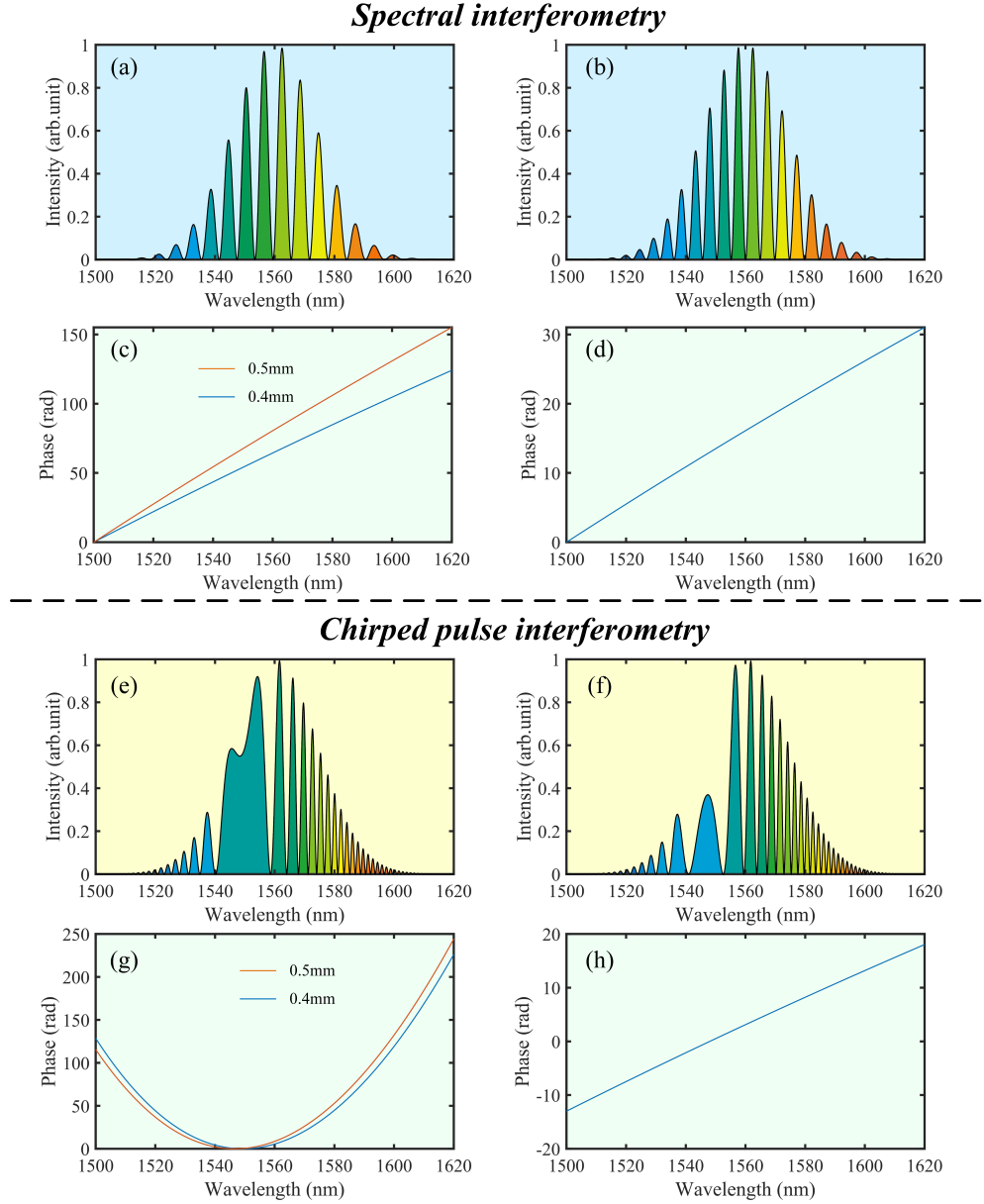}
    \caption{Comparison between the spectral interferometry and the chirped pulse interferometry. (a): Spectrogram in the spectral interferometry with 0.4 mm distance; (b): Spectrogram in the spectral interferometry with 0.5 mm distance. We find that the fringe will be modulated when the distance is not zero, and the modulation frequency increases with increasing the distance; (c): Unwrapped phases for the 0.4 mm and 0.5 mm distance, respectively. The unwrapped phase increases linearly with increasing the optical frequency; (d): Phase difference between the phases in (c). The distance can be determined by the phase slope; (e): Spectrogram in the chirped pulse interferometry with 0.4 mm distance; (f): Spectrogram in the chirped pulse interferometry with 0.5 mm distance. We find that the modulation frequency of the fringe is not constant any more. The position of the widest fringe is shifted when changing the distance; (g): Unwrapped phases for the 0.4 mm and 0.5 mm distance, respectively. The unwrapped phase is not linearly, but quadratically correlated with the optical frequency; (h): Phase difference between the phases in (g). We find that, the phase difference is still linearly related with the optical frequency, and the distance can be also determined by the phase slope.}
    \label{fig:3}
\end{figure*}

Therefore, the distances can be determined by:
\begin{equation} \label{eq:5}
L=\frac{1}{2}\cdot \frac{c}{n_g}\cdot \frac{d\varDelta \phi}{d\omega}
\end{equation}
where $c$ is the light speed in vacuum, and $n_g$ is the group refractive index of air. From Equation (\ref{eq:5}), we finally reach an expression related to the phase slope, exactly the same as spectral interferometry and dual-comb interferometry. This step is actually to measure the distance by time-of-flight method. As mentioned before, we try to efficiently remove the dead zones in the measurement path, and the spectrograms with one widest fringe shall be observed at any distances, as long as the dispersion is sufficiently large to cover the pulse-to-pulse length of the comb source (i.e., $c/(n_g{f_{rep}})$). In addition, the ambiguity range can be easily expanded to km level by slightly changing the repetition frequency. To make this idea more clearly, the comparison between the spectral interferometry and the chirped pulse interferometry with $1.1$×$10^{24}$ $\rm{rad/s^2}$ chirp rate is shown in {Figure \ref{fig:3}}. We find that, the method based on the phase slope can be universally used in the comb-based distance measurement, e.g., spectral interferometry, dual-comb interferometry, and chirped pulse interferometry.

In principle, the pulse width should be broadened to about 1/$f_{rep}$ to make sure that the pulses can always meet each other in space, to further remove the dead zones. {Please note that, the pulse width here does not mean the full width at half maximum, but means the total duration time of the pulse.} In this case, extremely long fiber is required, and the stabilization of such a long fiber is not easy (In fact, we used the 300 m long fiber to perform the distance measurement in our experiments, see APPENDIX). Although the long fiber can be stabilized by using cavity-stabilized laser, the system would become complicated and more expensive. In this work, we use a relatively short fiber, and change the repetition frequency of the frequency comb to ensure that the pulses can always overlap at arbitrary distances {\cite{50}}. Consequently, the measurement principle will be slightly updated. {The spectrogram corresponding to the reference mirror can be expressed as Equation (\ref{eq:2}) when the current repetition frequency is $f_{rep1}$. Here, the length difference between the reference pulses and the local oscillator can be calculated as:
\begin{equation} \label{eq:n6}
L_{ref}=\frac{1}{2}\cdot M_{ref}\cdot \frac{c}{n_g\cdot f_{rep1} }+\frac{1}{2}\cdot d_{ref}
\end{equation}
{where $M_{ref}$ is the pulse index difference between the pulses, which is related to the fiber length and the index. In Equation (\ref{eq:n6}), the first term at the right side is the integer part of the pulse-to-pulse interval, i.e., $c/(n_g \times f_{rep1})$, due to the pulsed nature of comb laser. The second term is the fractional part $d_{ref}$.}}

{In the case of the target mirror, the repetition frequency shall be increased to generate the spectrograms again (expressed as Equation (\ref{eq:3})), and the updated repetition frequency is $f_{rep2}$.
Consequently, we can calculate the length difference between the measurement pulses and the local oscillator as:
\begin{equation} \label{eq:n7}
L_{meas}=\frac{1}{2}\cdot M_{meas}\cdot \frac{c}{n_g\cdot f_{rep2} }+\frac{1}{2}\cdot d_{meas}
\end{equation}
{where $d_{meas}$ is the fractional part.}}

{Therefore, the measured distance (i.e., the difference between $L_{meas}$ and $L_{ref}$) can be calculated as:}
\begin{equation} \label{eq:6}
\begin{split}
L=&\frac{1}{2}\cdot M_{ref}\cdot \frac{c}{n_g\cdot f_{rep1} }-\frac{1}{2}\cdot M_{meas}\cdot \frac{c}{n_g\cdot f_{rep2} }+\frac{1}{2}\cdot \left(d_{meas}-d_{ref}\right)\\
=&\frac{1}{2}\cdot M_{ref}\cdot \frac{c}{n_g\cdot f_{rep1} }-\frac{1}{2}\cdot M_{meas}\cdot \frac{c}{n_g\cdot f_{rep2} }+\frac{1}{2}\cdot \frac{c}{n_g}\cdot \frac{d\varDelta \phi}{d\omega}
\end{split}
\end{equation}

We measure the distance by two steps. At the first step, the distances can be coarsely measured with a larger uncertainty based on Equation (\ref{eq:6}). Further, we use the synthetic wavelength interferometry to measure the distance. Since laser frequency comb is composed of numerous wavelengths, two wavelengths can be picked up to form a virtual synthetic wavelength, and the distances can be expressed as:
\begin{equation} \label{eq:7}
L=\frac{1}{2}\cdot \left( r+e \right) \cdot \frac{\varLambda}{{n_{g^{\mathrm{'}}}}}
\end{equation}
where $r$ is the integer part, $e$ is the fractional part of the synthetic wavelength, and ${n_{g^{\mathrm{'}}}}$ is the group refractive index. The synthetic wavelength $\varLambda$ can be calculated as:
\begin{equation} \label{eq:8}
\varLambda =\frac{\lambda _1\lambda _2}{\lambda _1-\lambda _2}
\end{equation}
where $\lambda_1$ and $\lambda_2$ are the chosen wavelengths. ${n_{g^{\mathrm{'}}}}$ can be written as:
\begin{equation} \label{eq:9}
{n_{g^{\mathrm{'}}}}=n_2-\lambda _2\frac{n_1-n_2}{\lambda _1-\lambda _2}
\end{equation}
where $n_1$ and $n_2$ are the corresponding phase refractive index of air. Please note that, the access condition from the first step to the second step is the measurement uncertainty of the first step is smaller than half the synthetic wavelength to obtain the right integer $r$. Finally, the distances can be finely determined by the two cascaded steps.

\section{Experimental Setup}
\begin{figure*}[htbp]
\centering
    \includegraphics[width=0.8\textwidth]{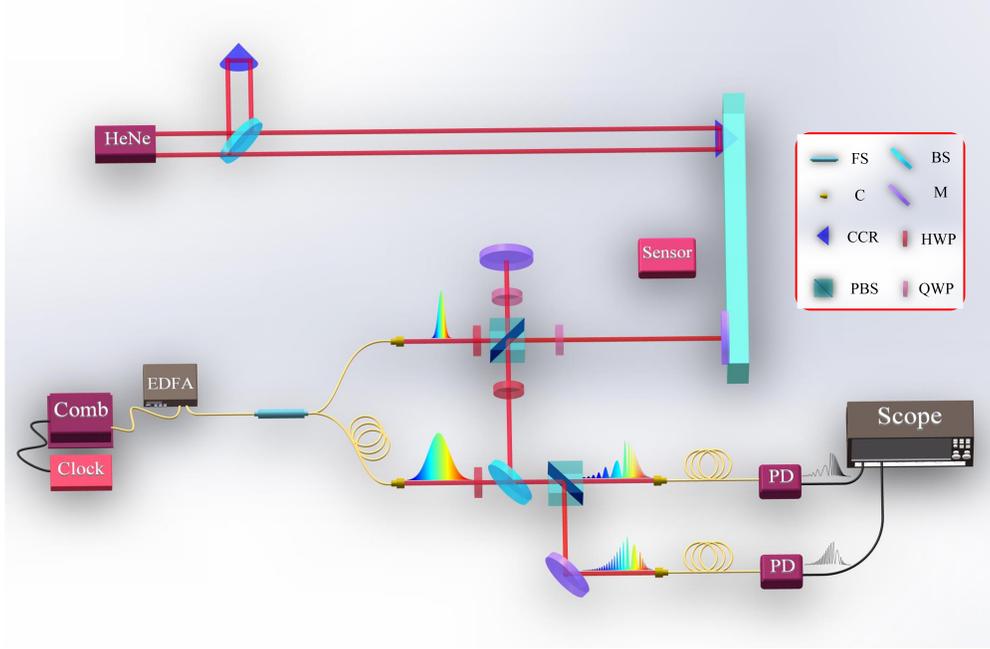}
    \caption{Experimental results of real-time chirped pulse interferometry for absolute distance measurement. EDFA: Er-doped fiber amplifier; PD: photodetector; FS: fiber splitter; BS: beam splitter; C: collimator; M: mirror; CCR: corner reflector; HWP: half wave plate; PBS: polarization beam splitter; QWP: quarter wave plate.}
    \label{fig:4}
\end{figure*}

\noindent
The experimental setup is shown in {Figure \ref{fig:4}}, which actually comprises of two interferometers, i.e., the frequency-comb interferometer and the HeNe interferometer. In the frequency-comb interferometer, the comb source (Menlosystems FC1500, 1560 nm center wavelength, about 250 MHz repetition frequency, 20 MHz carrier-envelope-offset frequency, 30 mW output power), locked to the Hydrogen maser (T4 Science iMaser 3000, 7.5×$10^{-14}$ @ 1s stability), is amplified by an Er-doped fiber amplifier to 130 mW, and then split into two parts. One part is broadened by a highly dispersive fiber (90 m dispersion compensation fiber), and works as the local oscillator. The other part goes into a Michelson interferometer, and serves as the signal source. The optical path difference between the reference arm (reflected by the reference mirror) and the measurement arm (reflected by the measurement mirror) is the distance $L$ to be determined. The target mirror is fixed on a PC-controlled carriage. The signal source and the local oscillator are combined at a beam splitter, and consequently we can obtain two spectrograms. One is formed by the reference pulse and the local oscillator, and the other corresponds to the measurement pulse and the local oscillator. These two spectrograms are split by a polarization beam splitter, and then injected into the highly dispersive fiber (2.5 km dispersion compensation fiber) to significantly stretch the spectrograms, respectively. Two fast photodetectors (Keyang KY-PRM-50G, 50 GHz bandwidth) are used to detect the spectrograms, and a high bandwidth oscilloscope (LeCroy WaveMaster 820Zi-B, 20 GHz bandwidth, 80 GS/s sampling rate) is exploited to measure and store the waveforms, which can be processed to extract the distance information. Please note that, a sinusoidal signal of about 250 MHz (corresponding to the current repetition frequency), which is locked to the Hydrogen maser, is used to trigger the oscilloscope.

To evaluate our results, a commercial distance meter (Renishaw XL80) is used to measure the same distance simultaneously in the HeNe interferometer, as depicted in Figure \ref{fig:4}. The target corner retroreflector of the commercial distance meter is fixed on the same carriage. Since the cw counting interferometer can only measure the incremental distances, we use the distance variations measured by our system to compare to the reference values. At the initial position, the reference distance meter is reset to zero, and we can measure an absolute distance of $L_0$. Then, the target mirror is moved far away from the beam splitter. The reference distance meter can measure the distance increment $\Delta$$L$, and meanwhile the updated absolute distance of $L_1$ can be determined by our system. Consequently, $(L_1- L_0)/2$ can be used to compare with the reference value of $\Delta$$L$. The environmental conditions are well controlled and measured, which can be used to calculate the air refractive index based on the empirical equations. To suppress the possible cosine error in long distance measurement, both the measurement beams of the HeNe interferometer and the frequency comb interferometer are carefully aligned to be parallel to the optical rail. The Abbe offset between the targets of the HeNe interferometer and the frequency-comb interferometer is less than 100 mm.
\section{Experimental Results}
\subsection{Mapping from the frequency domain to the time domain}
\begin{figure*}[htbp]
\centering
    \includegraphics[width=0.8\textwidth]{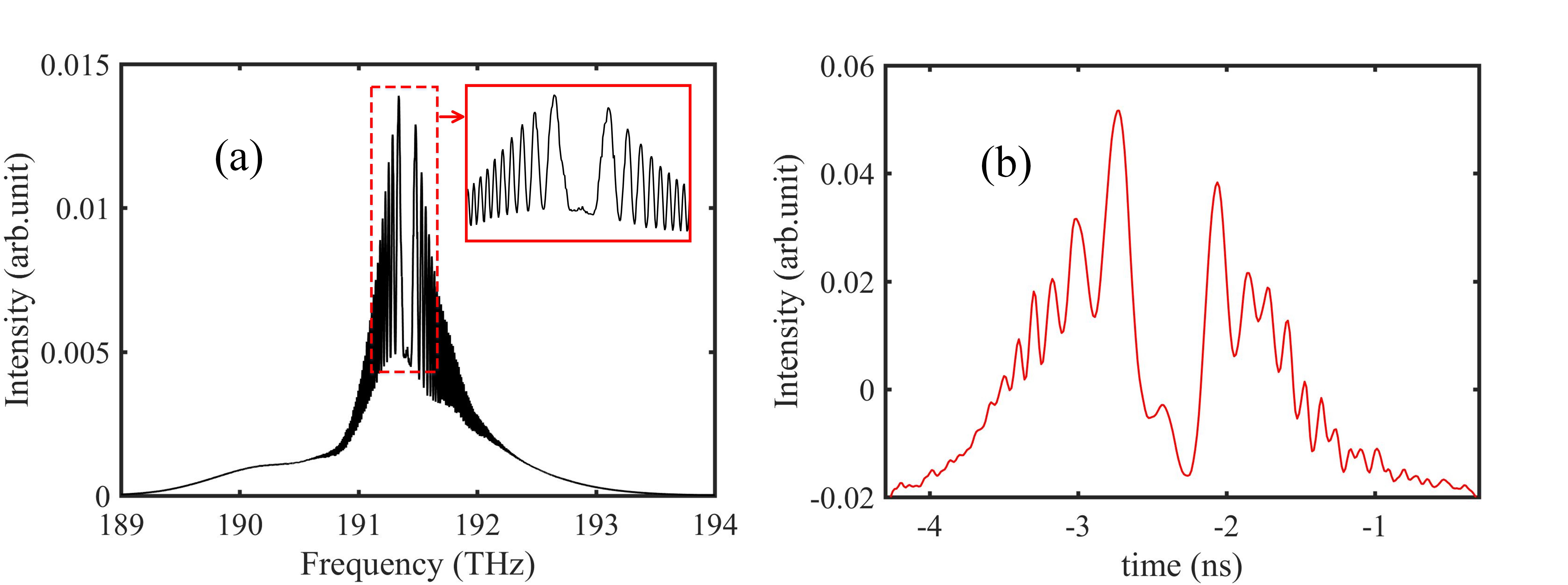}
    \caption{(a): Measurement results using optical spectrum analyzer; (b): Measurement results using the oscilloscope.}
    \label{fig:5}
\end{figure*}

\noindent
The precise mapping from the frequency domain to the time domain, i.e., from frequency to time is required in the practical measurement, which is determined by the parameters of the long dispersive fiber. Please see more details in APPENDIX. Unfortunately, the parameters provided by the fiber manufacturer are too rough to be used. In this step, we measure the spectrograms simultaneously by optical spectrum analyzer (YOKOGAWA AQ6370B) and the oscilloscope, and try to find the relation between the frequency and the time. {Figure \ref{fig:5}} shows a pair of measurement results. Figure \ref{fig:5}(a) shows the results measured by the optical spectrum analyzer. We find that the modulation frequency is not constant, and there is a widest fringe at about 191.4 THz. The fringe number is about 50. Figure \ref{fig:5}(b) indicates the results measured by the oscilloscope. Similar with that in Figure \ref{fig:5}(a), there is also a widest fringe at about -2.387 ns in the waveform. Limited by the resolution, we find that, 16 fringes can be clearly resolved.

\begin{figure}[htbp]
    \includegraphics[width=0.5\textwidth]{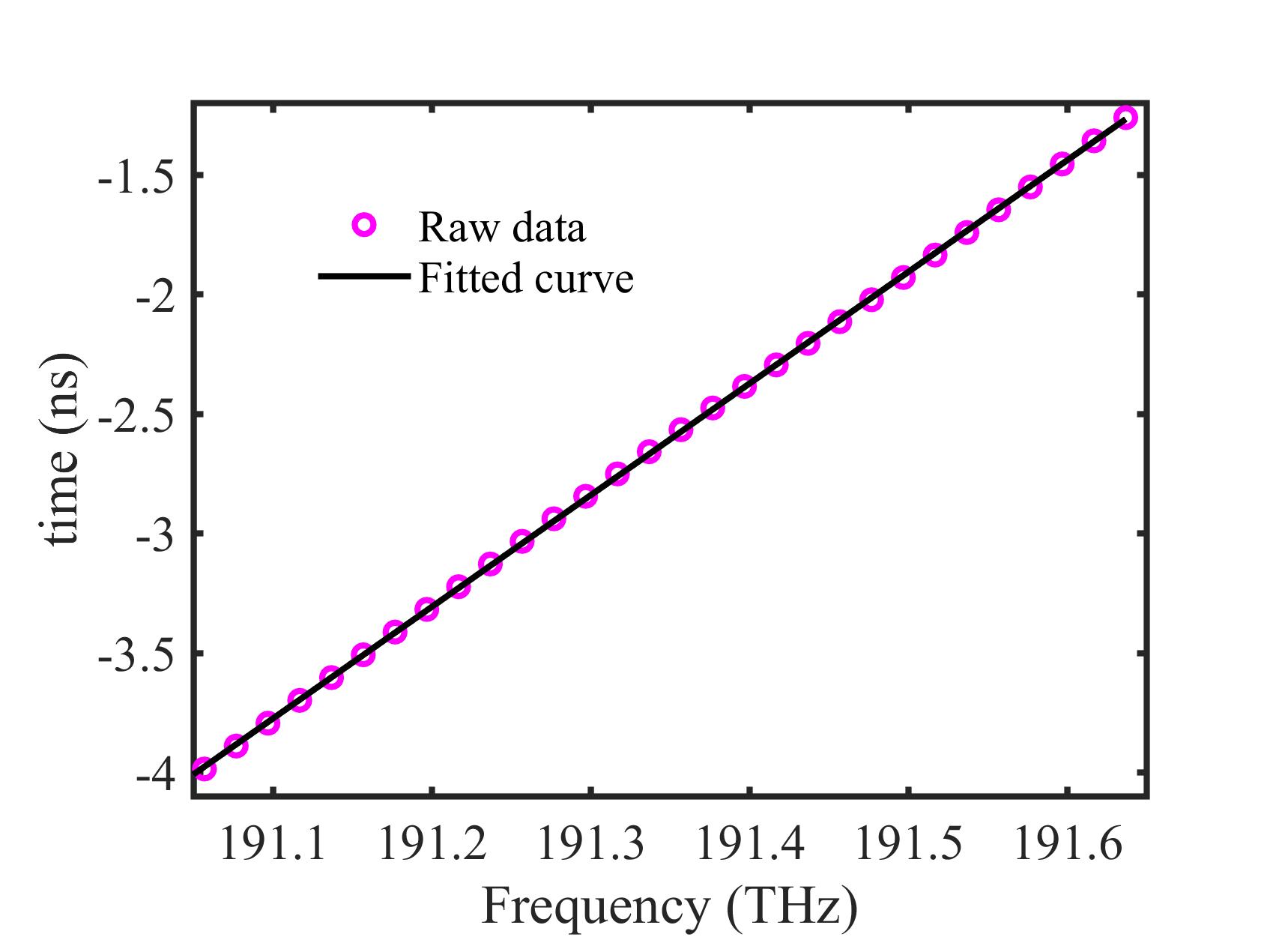}
    \caption{Relation between the time and the frequency. The pink solid circles show the raw data, and the black line indicates the fitted curve. We find that, time is linearly related to the frequency.}
    \label{fig:6}
\end{figure}

When the target mirror is moved far away, both the widest fringes on the OSA and the oscilloscope are shifted to the right side. We move the target mirror step by step, process both the data, and measure the exact position of the widest fringe. In the position of the widest fringe, the first order derivative of the unwrapped phase is equal to zero, i.e., the inflection point. The mapping from the frequency to the time is shown in {Figure \ref{fig:6}}, and can be fitted as $t=k\cdot{f}+b=4.669$×$10^{-21}$×$f-8.961$×$10^{-7}$, where $k$ is the slope and $b$ is the offset. $t$ and $f$ are the time (in s) and frequency (in Hz) values, respectively. In Figure \ref{fig:6}, the pink solid circles are the raw data, and the black line indicates the fitted curve. We find that, time is nearly linearly related to the frequency. Therefore, if using the time as the horizontal axis, Equation (\ref{eq:6}) can be updated to:

\begin{equation} \label{eq:10}
\begin{split}
L&=\frac{1}{2}\cdot M_{ref}\cdot \frac{c}{n_g\cdot f_{rep1} }-\frac{1}{2}\cdot M_{meas}\cdot \frac{c}{n_g\cdot f_{rep2} }+\frac{1}{2}\cdot \frac{c}{n_g}\cdot \frac{d\Delta \phi}{d\omega}\\
&=\frac{1}{2}\cdot M_{ref}\cdot \frac{c}{n_g\cdot f_{rep1} }-\frac{1}{2}\cdot M_{meas}\cdot \frac{c}{n_g\cdot f_{rep2} }\!+\!\frac{1}{2}\!\cdot\! \frac{c}{n_g}\!\cdot\! \frac{k}{2\pi}\!\cdot\! \frac{d\Delta \phi}{dt}
\end{split}
\end{equation}

\subsection*{Determination of the pulse index difference M}
\begin{figure*}[htbp]
\centering
    \includegraphics[width=0.7\textwidth]{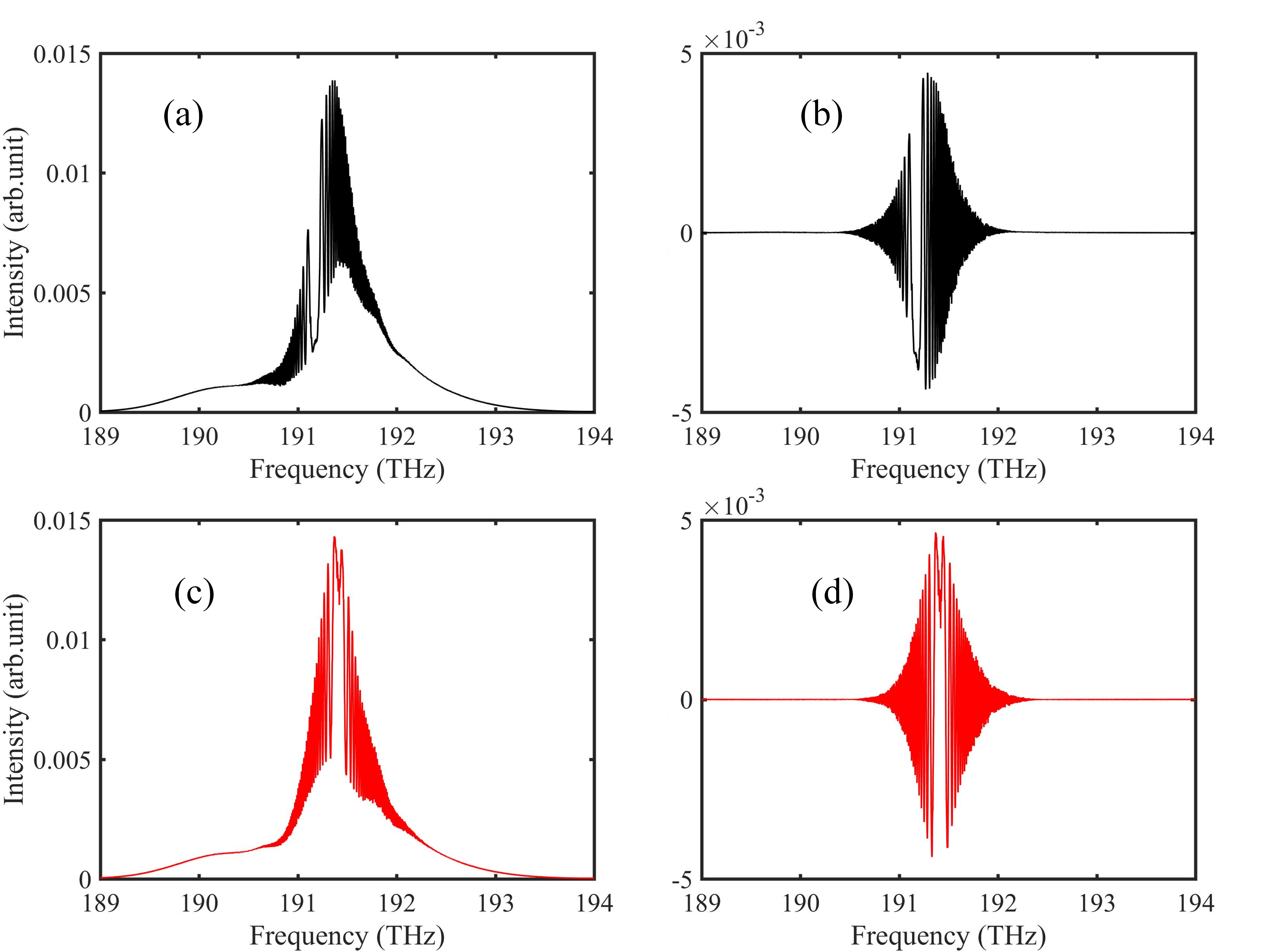}
    \caption{Spectrograms corresponding to the different repetition frequencies. (a): Spectrogram when the repetition frequency is 250.0300035 MHz. (b): AC part of the spectrogram in (a). (c): Spectrogram when the repetition frequency is 249.99951875 MHz. (d): AC part of the spectrogram in (c). The position of the widest fringe is shifted to the right side (higher optical frequency) when the repetition frequency decreases slightly. This is because the pulse-to-pulse interval increases when decreasing the repetition frequency.}
    \label{fig:7}
\end{figure*}

\noindent
As mentioned before, the position changes of the pulses from the local oscillator and the signal source are different due to the long fiber link. Considering the case of the reference pulses, the pulse index difference $M_{ref}$ can be measured by changing the repetition frequency, which is:
\begin{equation} \label{eq:11}
M_{ref}=round\left( \frac{\varDelta L\cdot f_{rep1}\cdot f_{rep2}}{c\cdot \varDelta f_{rep}} \right) 
\end{equation}
where $\Delta$$L$ is the change of the distance value due to the change of the repetition frequency. {Figure \ref{fig:7}} shows the experimental results for the determination of $M_{ref}$. When the repetition frequency is 250.0300035 MHz, the spectrogram is shown in Figure \ref{fig:7}(a), and we can find the widest fringe is located at about 191.2 THz. The widest fringe is shifted to about 191.4 THz as shown in Figure \ref{fig:7}(c) when the repetition frequency is changed to 249.99951875 MHz.

\begin{figure*}[htbp]
\centering
    \includegraphics[width=0.7\textwidth]{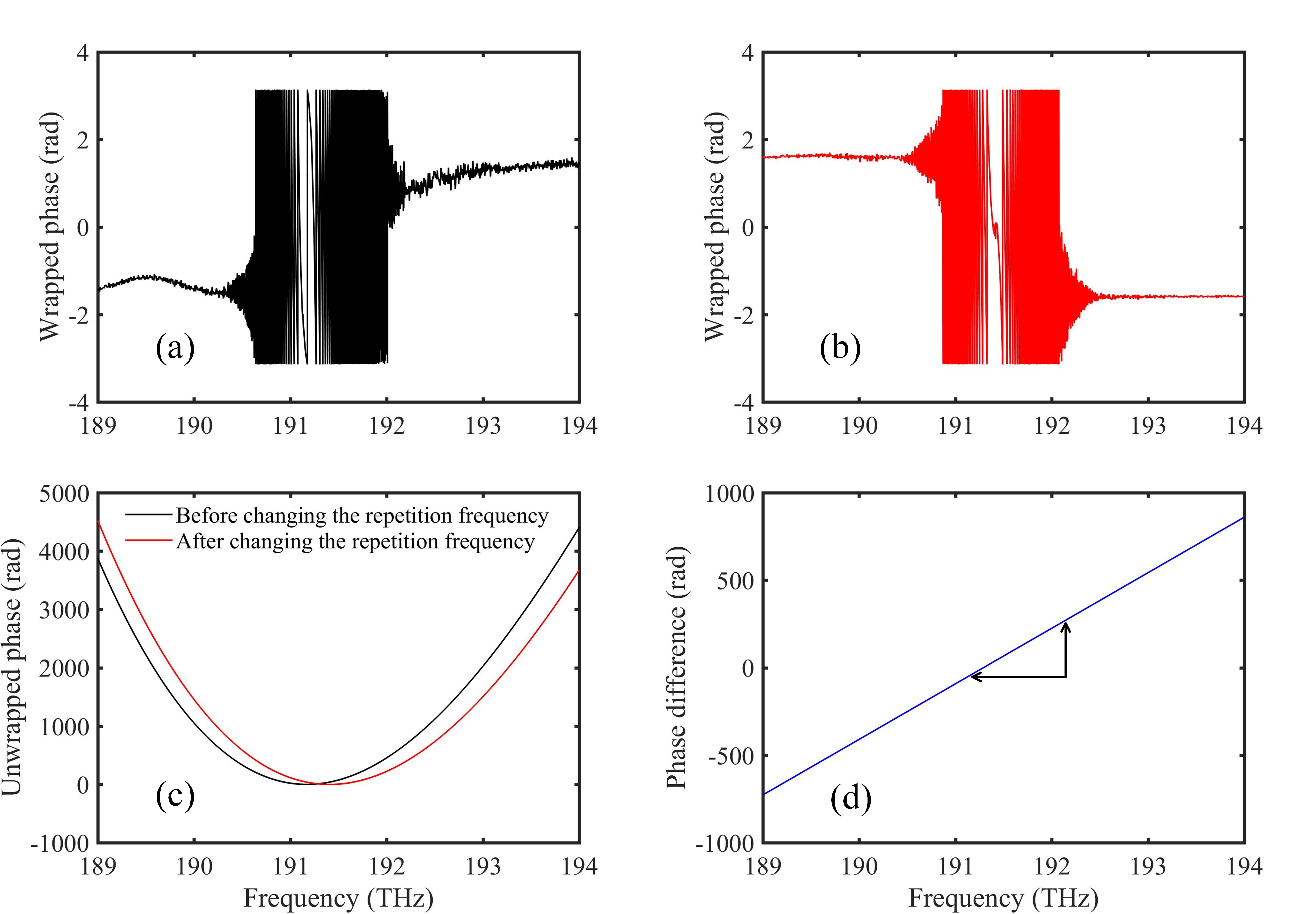}
    \caption{Data process of the spectrograms in the frequency domain. (a): Wrapped phase of the spectrogram in {Figure \ref{fig:7}}b; (b): Wrapped phase of the spectrogram in {Figure \ref{fig:7}}d; (c): Unwrapped phases. The black solid line represents the unwrapped phase before changing the repetition frequency, and the red solid line shows the phase after changing the repetition frequency. The unwrapped phase is shifted to the right side; (d): Phase difference between the two phases in {Figure \ref{fig:8}}c. We find a straight line whose slope can be used to determine the distances, despite that the unwrapped phase is not linearly related to the optical frequency.}
    \label{fig:8}
\end{figure*}

Here, we would like to demonstrate the data process in detail to obtain the distance information, which is shown in {Figure \ref{fig:8}}. Based on Hilbert transform, the wrapped phases of the signals shown in Figures \ref{fig:7}(b) and \ref{fig:7}(d) can be obtained as shown in Figures \ref{fig:8}(a) and \ref{fig:8}(b), respectively. We find the phases change the most slowly at the widest fringe. Figure \ref{fig:8}(c) indicates the unwrapped phases corresponding to the results in Figures \ref{fig:8}(a) and \ref{fig:8}(b). The black solid line represents the unwrapped phase when the repetition frequency is 250.0300035 MHz, and the red solid line shows the phase when the repetition frequency is 249.99951875 MHz. Figure \ref{fig:8}(d) shows the phase difference between the two phases in Figure \ref{fig:8}(c), which is a straight line. The distance can be measured from Equation (\ref{eq:5}), which turns out to be 15.06 mm. Inserting all the parameters into Equation (\ref{eq:11}), the $M_{ref}$ value can be determined to be 103, and  $M_{meas}$ can be also obtained  in this way. 
\subsection{Absolute distance measurement using real-time chirped pulse interferometry}
\begin{figure*}[htbp]
\centering
    \includegraphics[width=0.9\textwidth]{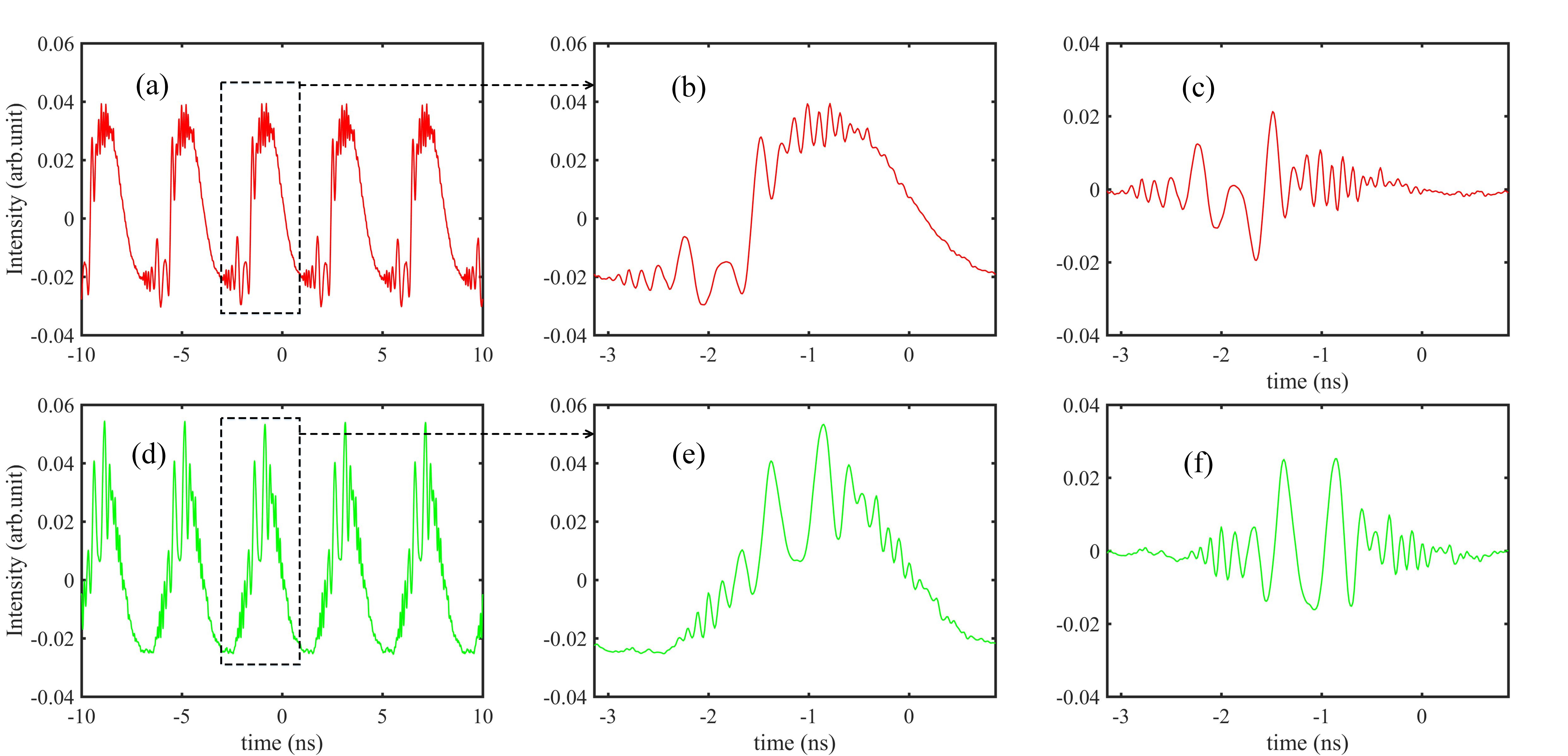}
    \caption{(a): Waveform with the target mirror at the initial position. (b): Waveform in one period of about 4 ns. (c): AC part of the waveform in (b). (d): Waveform with the target mirror shifted by 100 mm. (e): Waveform in one period of about 4 ns. (f): AC part of the waveform in (e).}
    \label{fig:9}
\end{figure*}

\noindent
In this subsection, we measure the distances by using real-time chirped pulse interferometry. {The environmental conditions are 22.5$^{\circ}$C, 98.8 kPa, and 63.1$\%$ humidity, and the group refractive index is 1.00026026} corrected by the Ciddor formula {\cite{51}}. {Figure \ref{fig:9}}(a) indicates the waveform at the initial position, and Figure \ref{fig:9}(c) shows the AC part in one period of about 4 ns. The repetition frequency is 250.0300035 MHz at this time. After the target mirror is shifted by about 100 mm far away, we change the repetition frequency to 249.6255035 MHz, so that the spectrograms can appear again on the oscilloscope. Figures \ref{fig:9}(d) and \ref{fig:9}(f) show the waveforms corresponding to the new position at about 100 mm. We use Hilbert transform to measure the phase of the waveforms shown in Figures \ref{fig:9}(c) and \ref{fig:9}(f), and the phases are then unwrapped. The slope of the phase difference can be used to determine the distances, just like the process shown in Figure \ref{fig:8}. The distance can be calculated based on Equation (\ref{eq:10}), which is 100.001 mm. Consequently, we use the synthetic wavelength interferometry to finely measure the distance.
	
\begin{figure*}[htbp]
\centering
    \includegraphics[width=0.9\textwidth]{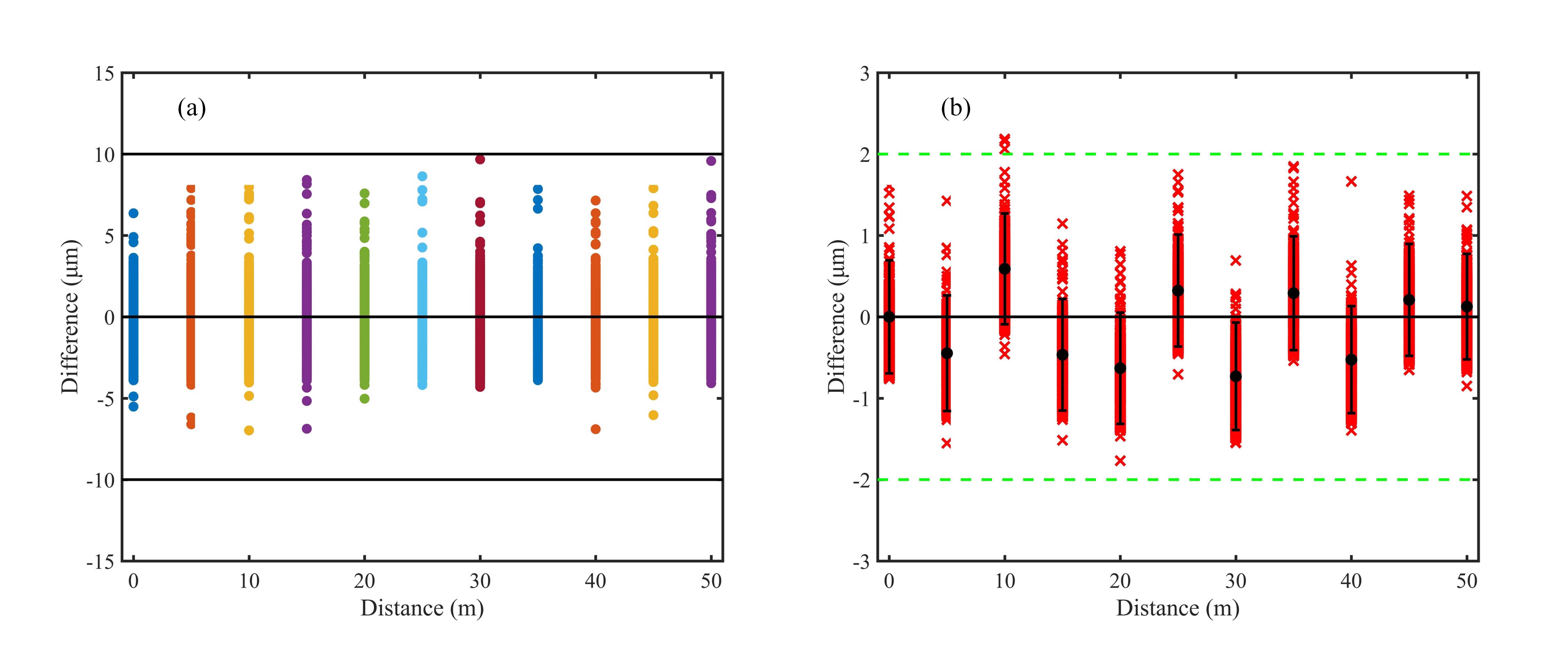}
    \caption{{Experimental results of the distance measurement. (a) Results of the coarse measurement; The colorful solid points show the scatters of each individual measurement; (b) Results of the fine measurement. The red x markers show the scatters of 1000 measurements. The black solid points indicate the average value, and the error bar shows twice the standard deviation.}}
    \label{fig:10}
\end{figure*}
The results of the distance measurement are shown in {Figure \ref{fig:10}}, and we measure the distances by two steps. The horizontal axis is the reference distances, and the vertical axis is the difference between our measurements and the reference values. Figure 10a indicates the results of the coarse measurements. We measure the distances for 1000 times (for just 4 $\upmu$s in fact) at each position, and the colorful solid points are the scatters of each individual measurements. The measuring time for each measurement is about 4 ns, corresponding to the repetition frequency of the laser frequency comb. We find that, the measurement uncertainty can be below ±10 $\upmu$m at this step. In the second step, i.e., the fine measurement, the synthetic wavelength should be larger than 2×20 $\upmu$m. We choose the wavelengths of 1560 nm and 1570 nm to generate the synthetic wavelength. Correspondingly, the group refractive index is 1.00026027. The synthetic wavelength is therefore 244.9 $\upmu$m, \textgreater 2×20 $\upmu$m. Based on Equation (\ref{eq:9}), the results of the fine measurement are shown in Figure \ref{fig:10}(b). The scatters of 1000 measurements are indicated by the red x markers. The black solid points represent the average value of the 1000 measurements, and the error bar shows twice the standard deviation. The green dashed lines show the limit of the measurement uncertainty, which is improved to be below ±2 $\upmu$m.

\begin{figure}[htbp]
    \includegraphics[width=0.5\textwidth]{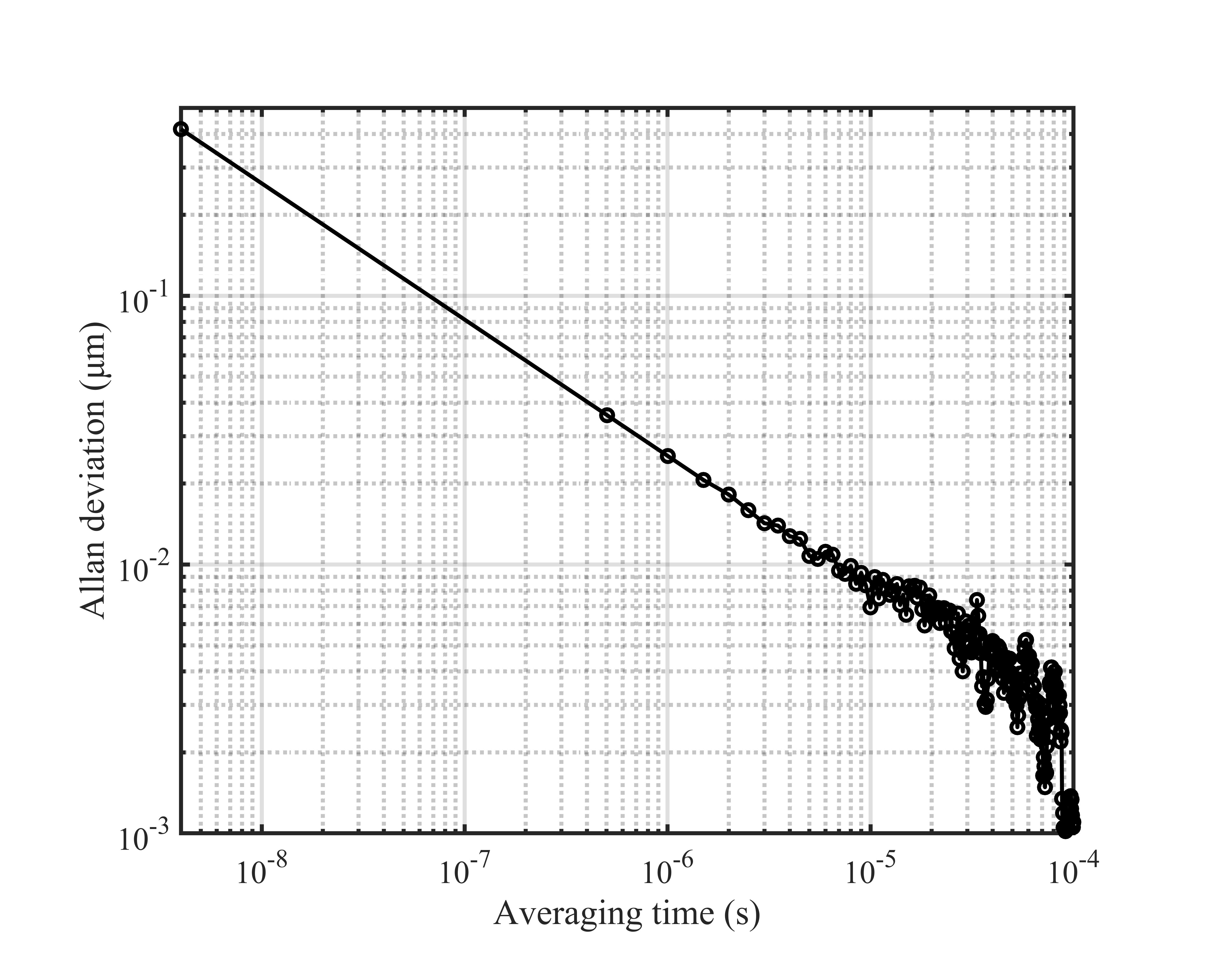}
    \caption{Allan deviation at different averaging time.}
    \label{fig:11}
\end{figure}

{We carry out long-term experiments for 800 $\upmu$s with 200000 measurements, to examine the precision limit. {Figure \ref{fig:11}} shows the Allan deviation at shows the Allan deviation at 50 m distance. The Allan deviation is 0.4 $\upmu$m at averaging time of 4 ns, 25 nm at 1 $\upmu$s, and can achieve 2 nm at 100 $\upmu$s averaging time. These results show that nanometer-level precision can be realized.}

\subsection{Ultra-fast distance measurement}
\begin{figure*}[htbp]
\centering
    \includegraphics[width=0.8\textwidth]{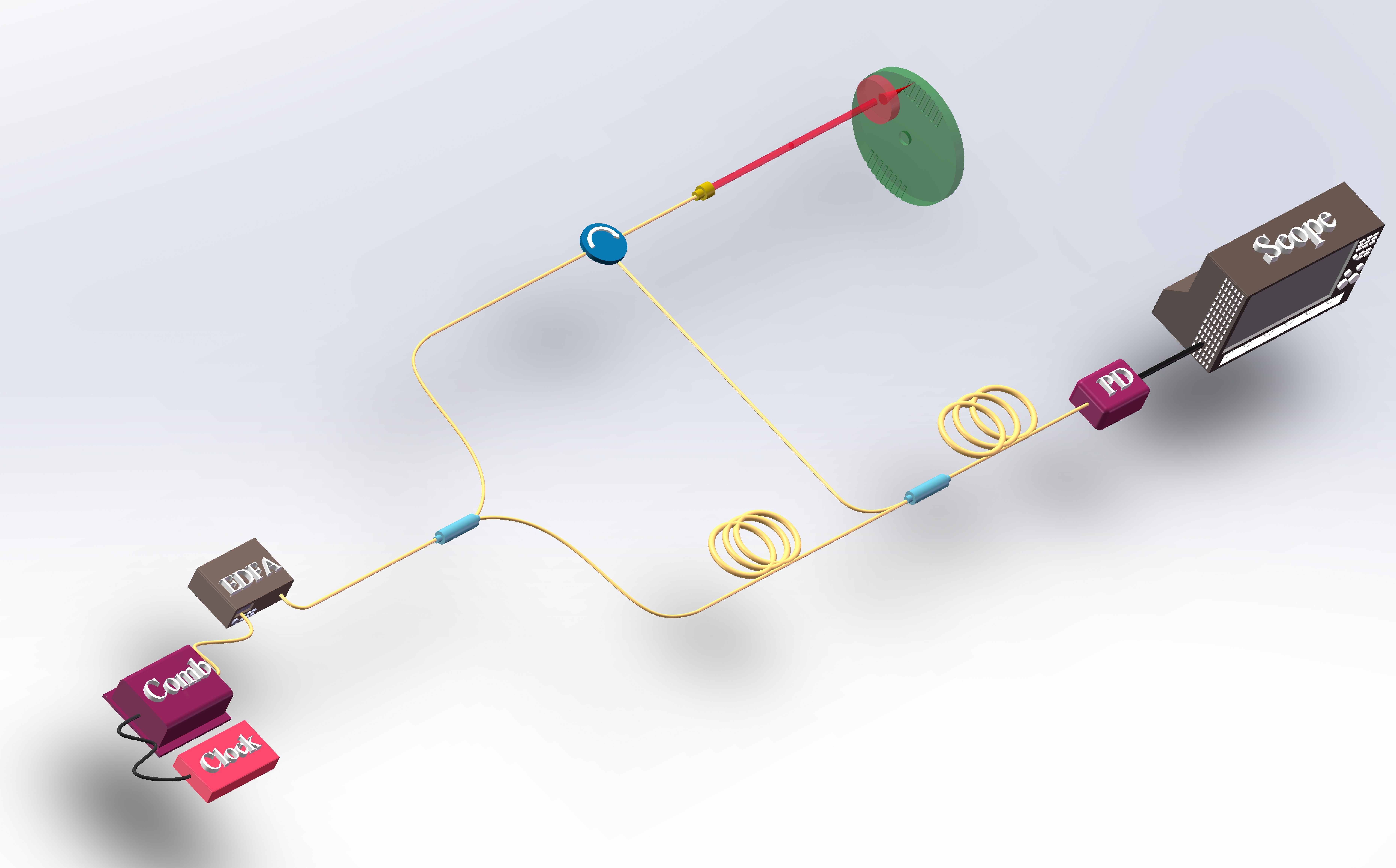}
    \caption{Experimental setup of ultrafast distance measurement. The target is a spinning disk with grooves of different depths.}
    \label{fig:12}
\end{figure*}

\begin{figure}[htbp]
    \includegraphics[width=0.5\textwidth]{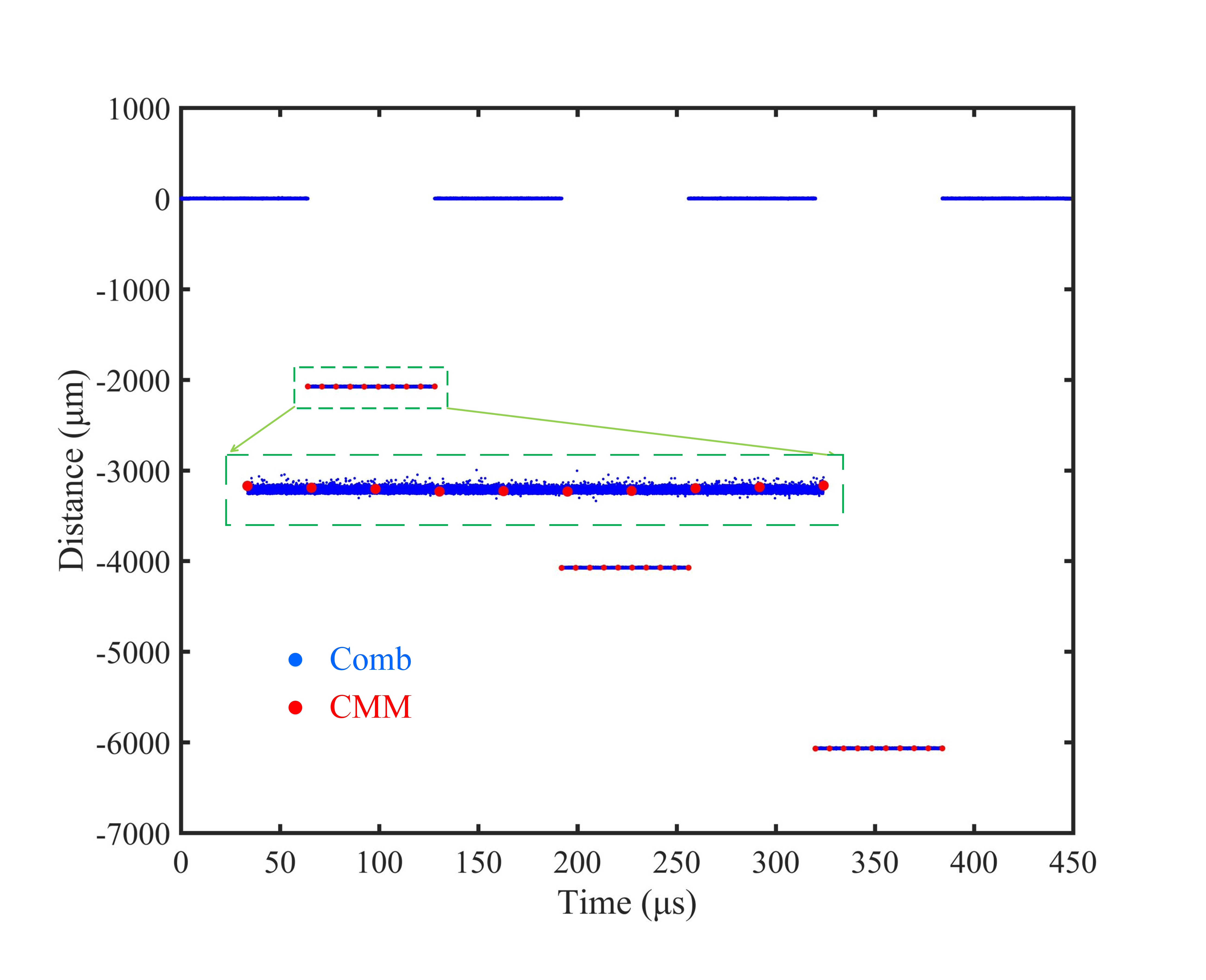}
    \caption{Experimental results of ultrafast distance measurement. The blue points represent the results using real-time chirped pulse interferometry, and the red solid points indicate the results measured by a coordinate measuring machine.}
    \label{fig:13}
\end{figure}

\noindent
Real-time chirped pulse interferometry is able to measure the distances with ultrafast speed, which is only limited by the repetition frequency of the laser source. The experimental configuration is shown in {Figure \ref{fig:12}}. The measurement beam is injected into the circular, and focused onto the spinning disk. After reflected by the spinning disk, the measurement beam is combined with the local oscillator, stretched by a highly dispersive fiber, and finally detected by a fast photodetector. The radius of the spinning disk is about 15 cm, and the rotating speed is about 10000 rpm. Therefore, the line speed of the edge is about 157 m/s. Several grooves with different depths are designed on the surface of the disk. Please note that, the length of the dispersive fiber here is about 20 m, since the groove depths are not very large in the experiments. The measurement results of these grooves are shown in {Figure \ref{fig:13}}. The blue points represent the results using real-time chirped pulse interferometry, and the red solid points indicate the results measured by a coordinate measuring machine (CMM, Leica PMM-Ultra). The difference between the two measurements is below ±2 $\upmu$m.
\section{Uncertainty evaluation and discussion}
\noindent
The final measurement results are based on Equation (\ref{eq:7}), which involves the parameters of the integer part $r$, the fractional part $e$, the synthetic wavelength $\varLambda$, and the group refractive index  ${n_{g^{\mathrm{'}}}}$. The integer part $r$ can be precisely determined with uncertainty better than 0.5, and has no contributions to the measurement uncertainty. The uncertainty $u_L$ can be calculated by:
\begin{equation} \label{eq:12}
u_L=\sqrt{\left( \frac{L}{\varLambda}\cdot u_{\varLambda} \right) ^2+\left( \frac{L}{n_g}\cdot u_{{n_{g^\mathrm{'}}}} \right) ^2+\left( \frac{1}{2}\cdot \frac{L}{n_g}\cdot u_e \right) ^2}
\end{equation}

In Equation (\ref{eq:12}), the first term at the right side is related to the synthetic wavelength $\varLambda$. In our experiments, laser frequency comb is referenced to the Hydrogen maser with 7.5×$10^{-14}$ stability at 1 s, and each frequency marker holds the same frequency stability. We use a well-calibrated optical spectrum analyser to measure the wavelengths. Therefore, it is reasonable and sufficient that the uncertainty due to the clock stability can be evaluated to be $10^{-13}L$. The precise mapping from the frequency domain to the time domain can also make contribution to the first term, which is related to the stability of the long fiber and the stability of the clock source for the oscilloscope. The long fibers used in the detection unit are well vibration-isolated and temperature controlled, and the temperature drift can achieve 1 mK over 2 days. In addition, the oscilloscope is locked to the Hydrogen maser. We measure the mapping for 2 days continuously, and for 1565 nm the stability of the corresponding time can reach 0.013 fs, which corresponds to 4.1×$10^{-9}L$ uncertainty. The second term is related to the group refractive index ${n_{g^{\mathrm{'}}}}$. We measure the air refractive index using Ciddor formula based on the environmental sensor network. The inherent uncertainty of the empirical formula is 2×$10^{-8}$, and the measurement of the environmental conditions also plays an important role. With consideration of the uncertainty of the sensors, the environmental stability, and the environmental inhomogeneity, the uncertainties of the temperature, air pressure, and humidity are 28 mK, 12 Pa, and 1.6$\%$, which respectively correspond to 2.6×$10^{-8}$, 3.1×$10^{-8}$, and 1.7×$10^{-8}$ uncertainty of air refractive index. The combined uncertainty of the air refractive index can thus reach 9.7×$10^{-8}$ (k=2), which is 9.7×$10^{-8}L$ for the distance uncertainty. The third term is due to the measurement of the fractional part, which can be estimated by the standard deviation. At each position, we measure the distance by 1000 times with ultrafast measurement speed, and the standard deviation is below 0.6 $\upmu$m. We use twice the standard deviation to evaluate this part, which is less than 1.2 $\upmu$m. Since the frequency comb laser and the reference distance meter do not share the same target mirror, Abbe error could exist. The Abbe error is always below 2 $\upmu$m along the 50 m travel, corresponding to a 20 $\upmu$rad yaw error of the rail. In addition, the stability of the translation stage is below 0.2 $\upmu$m over 2 hours. Finally, the combined uncertainty of the distance measurement is $\rm[(2.4\upmu{m})^2+(9.7\times10^{-8}\cdot{L})^2]^{1/2}$, which can well cover the results in Figure \ref{fig:10}(b). The uncertainty evaluation is summarized in {Table \ref{tab:1}}.
\begin{table*}[htbp]
    \caption{Uncertainty budget of the absolute distance measurement.}    
    \begin{tabular}{ccc}
            \hline
           Sources of the measurement uncertainty & Value \\  
            \hline
            Due to the synthetic wavelength& $10^{-13}\cdot{L}$ \\ 
            Due to the air refractive index & $9.7\times{10^{-8}}\cdot{L}$
            \\
            Temperature uncertainty&$28 \rm mK$
            \\
            Air pressure uncertainty&$12 \rm Pa$
            \\
            Humidity uncertainty&$1.6\%$
            \\
            Due to the fractional part&$1.2$ $\upmu$$\rm m$
            \\
            Abbe error&$2$ $\upmu$$\rm m$
            \\
            Due to the mapping from the frequency domain to the time domain&$4.1\times{10^{-9}}\cdot{L}$
            \\
            Due to the long optical rail&$0.2$ $\upmu$$\rm m$
            \\
             \hline
            Combined uncertainty (k = 2)&$[(2.4 \upmu\rm{m})^{2}+(9.7\times10^{-8}\cdot{L})^2]^{1/2}$
            \\
             \hline
            \end{tabular}
    \label{tab:1}
\end{table*}

  Further, we would like to give an analysis about the measurement resolution, which is very important in the practical distance measurement. In general and basically, the measurement resolution is related to the linewidth of the optical source, the phase resolution of the measurement system, and the short-term stability of the environment. Right now, the optical source can be locked to a reference cavity or an atomic clock (the Hydrogen maser in this work), and the resulting linewidth of the optical source can reach Hz or even sub Hz {\cite{52}}, corresponding to much better than pm resolution. Here, the phase resolution of the phasemeter could make the main contribution. In our work, we use the oscilloscope and the post data process to measure the phases. The oscilloscope is locked to the Hydrogen maser, and the digital noise and the ADC (AD card) noise can be neglected. In this case, we can use the value provided by the manufacturer, and the phase resolution can reach 0.01 degree. Therefore, the length resolution can achieve 0.01/360×244.9$\upmu$m (synthetic wavelength), and that is 6.8 nm. We find, smaller synthetic wavelength can lead to better length resolution. If we use the single wavelength to measure the distances, the measurement resolution can be expected to be better. However, generally speaking, coherent averaging is needed to reach such a high performance for high-speed measurement. Of course, there are the other methods (e.g., optical method) for the phase detection. Finally, the stability of the environment could affect the measurement resolution. In the case of the long distance measurement, it is difficult to maintain the environment homogeneity and stability. Despite this, the nanometric resolution has been reported in the previous reports {\cite{29,30}}.

  In discussion, first the distances can be precisely determined via the spectrograms shown in Figure \ref{fig:7} actually. In this case, the mapping from the frequency domain to the time domain is not required. Therefore, the calibration step in Figure \ref{fig:6} is not needed any more. In general, the schemes based on the chirped pulse interferometry involve one calibration step, so that the distance can be measured via the position of the widest fringes. We would like to show that, the distance can be also measured by the slope of the phase difference in the chirped pulse interferometry, even if the phase of one spectrogram is not linear, but quadratic. The main advantage of the real-time chirped pulse interferometry is that the measurement speed can be improved to be record high, i.e., 4 ns per measurement. Second, in our experiments, one long fiber link is involved to stretch the femtosecond pulses in the local oscillator. The active stabilization of the fiber in the local oscillator is not strictly required. The two spectrograms used to measure the distances always share the same fiber in local oscillator. Therefore, the fiber fluctuations can be cancelled out in real time. Finally, we use 90-m dispersion compensation fiber in the local oscillator, and therefore the repetition frequency should be changed so as to remove the dead zones. In fact, at the very beginning we used 300-m dispersion compensation fiber in the local oscillator, and the adjustment of the repetition frequency is not needed in that case. The problem is that less fringes can be obtained on the oscilloscope if with such a long fiber because the sampling rate of the oscilloscope is limited. Another is that the stabilization of such a long fiber is not easy, and our experimental results also show that shorter fiber can lead to a better performance of distance measurement. Please find more details in APPENDIX. In addition, we use a fully-stabilized frequency comb in our experiments, and the locking circuits indeed make the whole system bulky and expensive. We consider that, the stabilization of the repetition frequency is needed. If the repetition frequency is free, the pulse-to-pulse interval will be free, and the spectrograms can not be sufficiently stable. The position of the widest fringe will fluctuate, leading to the performance degradation. The $f_{ceo}$ locking could not be strictly required, because the spectrograms in chirped pulse interferometry is related to the temporal positions of the wave packets. If the optical phase of one wavelength is used in the distance measurement, the $f_{ceo}$ locking is also suggested. We would like to discuss the implementation of the soliton microcombs in the real-time chirped pulse interferometry. The repetition frequency of microcomb is often high at tens of GHz because of the small footprint. If with 10 GHz repetition frequency, the pulse period is 100 ps. Consequently, the sample number is 8 with 80 GSa/s sampling rate, which makes the observation of the spectrograms inaccessible. Therefore, to observe the birth of microcomb solitons is rather difficult.

\section{Conclusion}
\noindent
In this work, we describe a method based on real-time chirped pulse interferometry, capable of ultrafast, precise absolute distance measurement without dead zones along the measurement path. The local oscillator is strongly chirped by a highly dispersive fiber, so that the measurement pulses can always leave footprints on the spectrograms to remove the dead zones. Due to the strong chirp, the spectral phase of one spectrogram is not linearly correlated to the optical frequency. In spite of this, the phase difference between two spectrograms is still a straight line, whose slope can be used to measure the distances. Next, virtual synthetic wavelength interferometry is exploited to finely determine the distances. In contrast to the previous reports {\cite{47,48}}, the step to calibrate the position of the widest fringe is not required. We develop a real-time optical spectrum analyzer based on dispersive Fourier transform. Because of the existence of the widest fringe, we can observe the spectrograms on the oscilloscope at any distances, which can relax the restriction of the time resolution of the scope. We carry out the experiments in the metrology lab, and the comparison to the reference distance meter shows that the difference can be well below ±2 $\upmu$m. There is no dead zones in the measurement path. This level of measurement uncertainty can satisfy most of the applications actually. A spinning disk with about 150 m/s line speed can be imaged due to the record high measurement speed. Our method offers a unique and comprehensive combination of non-dead zones, high speed (4 ns per measurement), high precision (nanometer level), large ambiguity range (up to km level by slightly changing the repetition frequency), and {with only one comb source}.

To date, the technique of frequency combs is becoming mature, and has seen a wealth of applications in both science and technology. However, most of the applications based on the combs are only at the level of laboratory demonstration, and there is still a long way to go to realize the practical applications in the industry. In fact, the limitation of the comb source itself has been solved, and portable and cost-efficient combs are available. Scientists have tried hard to develop the frequency-comb instruments, e.g., distance meter, 3D scanner, space-borne frequency combs, and spectrometer, etc {\cite{2}}. But, the market share is not comparable to the traditional metrology instruments. We focus on the technique of distance metrology, and precision measurement of the physical quantity related to the distance. Considering the future field measurement using laser frequency combs, frequency comb Lidar (laser-based light detection and ranging) using the technique demonstrated in this work is feasible, which can realize 3D measurement at long range with the weak light detection {\cite{53}}. If out of the lab, the air refractive index is going to make great contribution to the final uncertainty. The two-color or three-color method can be used to correct the air refractive index in real time in this case {\cite{54,55}}. Our work could provide a solution for the practical applications using frequency combs in future.

\begin{appendices}
\setcounter{equation}{0}
\renewcommand{\theequation}{A\arabic{equation}}
\setcounter{figure}{0}
\renewcommand{\thefigure}{A\arabic{figure}}
\setcounter{table}{0}
\renewcommand{\thetable}{A\arabic{table}}

\section{Pulse broadening using the dispersion compensation fiber}\label{A1}
\noindent In this section, we would like to give a theoretical description of the pulse broadening in the chirped pulse interferometry in detail. The fs-class pulses will be broadened when travelling in a dispersive media, which is the dispersion compensation fiber DCF in this work. If the spectrum of the frequency comb can be expressed by $E(\omega)$, the electrical field of one pulse can be thus calculated based on the inverse Fourier transform as:
\begin{equation} \label{eq:s1}
E\left( t \right) =\frac{1}{2\pi}\int_{-\infty}^{+\infty}{E\left( \omega \right)}\cdot e^{i\omega t}d\omega 
\end{equation}  
where $\omega$ is the angular frequency, and the initial phase is zero. After propagating through the DCF, an additional phase will be introduced for each mode due to the length and the refractive index (i.e., the dispersion) of the fiber. This additional phase $\psi (\omega)$ can be expressed as:
\begin{equation} \label{eq:s2}
\psi \left( \omega \right) =\sum\nolimits_{m=1}^{\infty}{\frac{\beta _m}{m!}}\left( \omega -\omega _c \right) ^mz=\beta _1\left( \omega -\omega _c \right) z+\frac{\beta _2}{2}\left( \omega -\omega _c \right) ^2z+\frac{\beta _3}{6}\left( \omega -\omega _c \right) ^3z...
\end{equation}  
where $\beta_m$ is the coefficient of the group velocity dispersion GVD at different orders, $m$ is an integer, $\omega_c$ is the center frequency, and $z$ is the fiber length. Consequently, the spectrum of the laser source can be updated to:
\begin{equation} \label{eq:s3}
E_1\left( \omega \right) =E\left( \omega \right) \cdot e^{i\psi \left( \omega \right)}
\end{equation}  

Therefore, the electrical field of the pulse can be given by
\begin{equation} \label{eq:s4}
E\left( t \right) =\frac{1}{2\pi}\int_{-\infty}^{+\infty}{E\left( \omega \right)}\cdot e^{i\psi \left( \omega \right)}\cdot e^{i\omega t}d\omega 
\end{equation}  

Inserting Eq. (\ref{eq:s2}) into Eq. (\ref{eq:s4}), we can reach
\begin{equation} \label{eq:s5}
\begin{split}
E\left( t \right) &=\frac{1}{2\pi}\int_{-\infty}^{+\infty}{E\left( \omega \right)}\cdot exp\!\:\left( i\sum\nolimits_{m=1}^{\infty}{\frac{\beta _m}{m!}}\left( \omega -\omega _c \right) ^mz+i\omega t \right) d\omega \\
&=\frac{1}{2\pi}\int_{\omega _c-\Delta \omega /2}^{\omega _c+\Delta \omega /2}{E}\left( \omega \right) \cdot exp\left( i\sum_{m=1}^{\infty}{\frac{\beta _m}{m!}}\left( \omega -\omega _c \right) ^mz+i\omega t \right) d\omega 
\end{split}
\end{equation}  
where $\Delta{\omega}$ is the spectral width. Here, we finally have got an expression of the electrical field after the laser pulses travel through the DCF. We find, larger $\Delta{\omega}$, larger $\beta_m$, and larger $z$ can result in a larger pulse width.

\begin{figure}[!h]
\centering
    \includegraphics[width=0.6\textwidth]{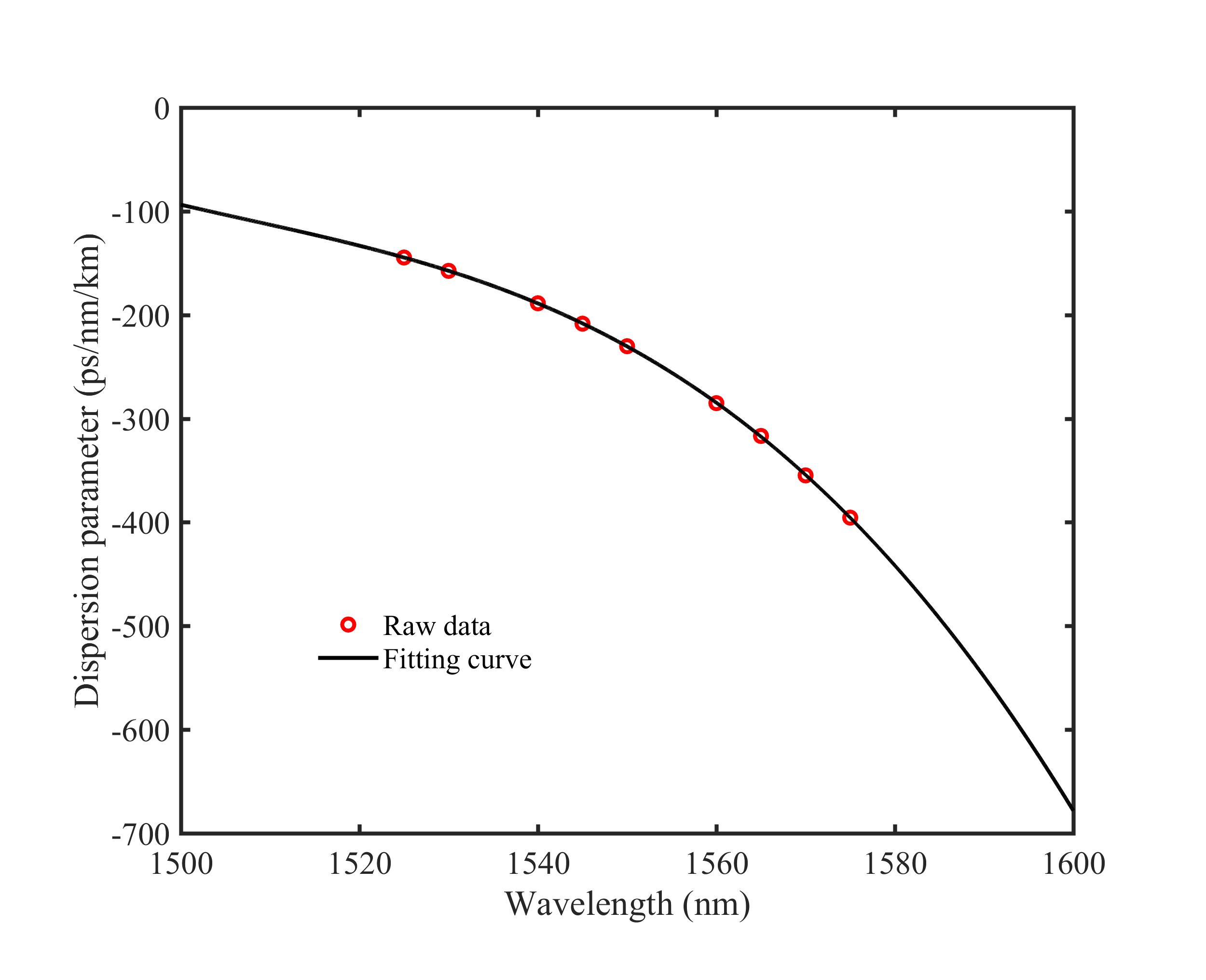}
    \caption{Dispersion parameters provided by the fiber manufacturer. Red circles show the raw data, and the black line indicates the fitting curve.}
    \label{fig:s1}
\end{figure}

We carry out the simulations to guide the experimental configurations by using the parameters provided by the fiber manufacturer. Fig. \ref{fig:s1} shows the dispersion parameters with unit of ps/ns/km, and the coefficients of GVD can be obtained accordingly, as shown in Table \ref{tab:s1}.
 
\begin{figure}[!h]
\centering
    \includegraphics[width=0.9\textwidth]{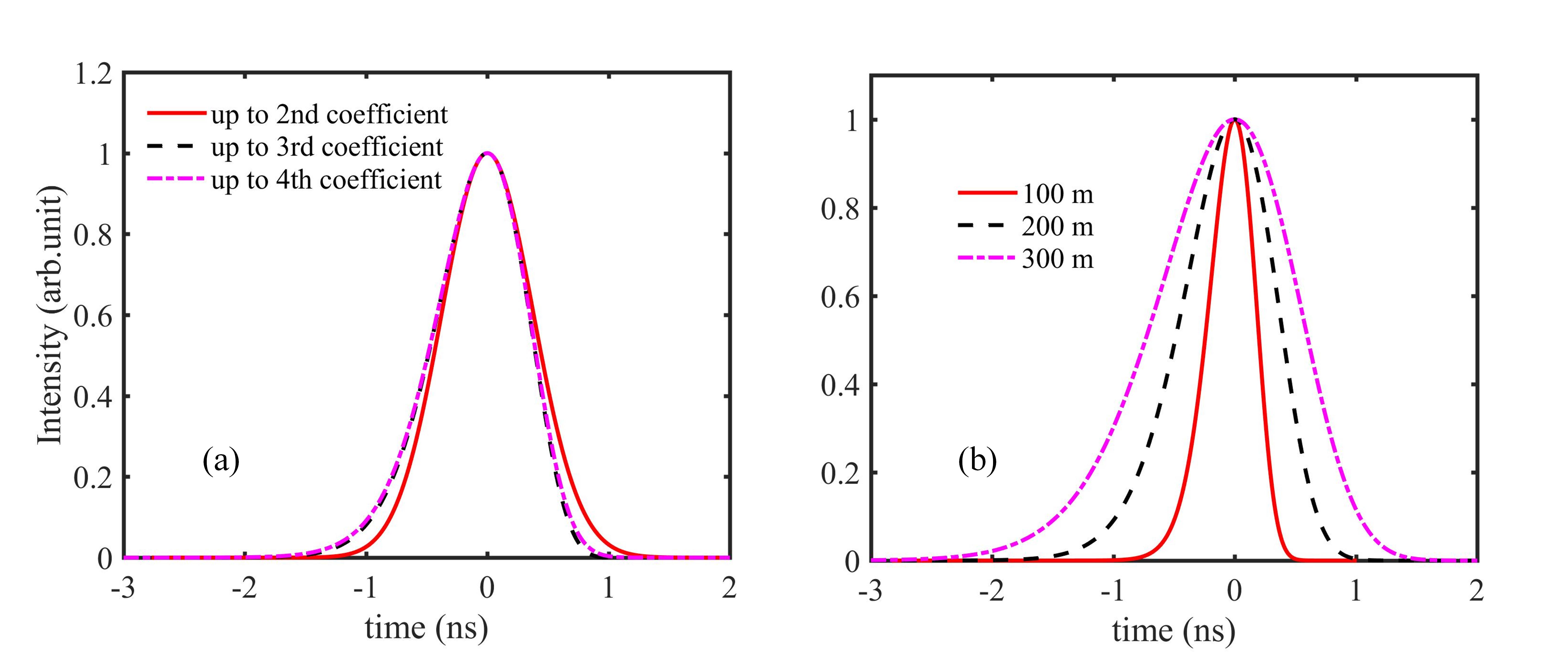}
    \caption{(a): Pulse shapes with different orders of dispersion coefficients, when the fiber length is 300 m;(b): Pulse shapes with different fiber lengths, while the dispersion involved is up to the 4th coefficient of GVD.}
    \label{fig:s2}
\end{figure}

In our experiments, the center wavelength of the comb source is about 1560 nm, and the spectral width is about 2 THz. Based on Eq. (\ref{eq:s5}) and using the parameters in Table \ref{tab:s1}, we can obtain the pulse shape after the pulse propagates through the long fiber. Fig. \ref{fig:s2}(a) indicates the pulse shapes with different orders of dispersion coefficients, while the fiber length is 300 m. We find, the pulse shape is well symmetrical with up to the $\rm 2^{nd}$ dispersion coefficients, i.e., the maximum value of m is 2 in Eq. (\ref{eq:s5}). When the order of the dispersion coefficients increases, the pulse shape is not symmetrical any more, and keeps almost the same. In order to completely remove the dead zones in the measurement path, the pulse width should be close to 4 ns, i.e., 1/250MHz. As a consequence, we examine the pulse width with the different fiber lengths, as shown in Fig. \ref{fig:s2}(b). It is clear that the pulse width is becoming larger when increasing the fiber length. We find that, the pulse width is about 4 ns when the fiber length is 300 m.
\begin{table}[!h]
\centering
    \caption{Coefficients of GVD at different orders.}    
    \begin{tabular}{ccc}
            \hline
           $\beta_2$ & $\rm29ps^2/m$ \\  
            \hline
             $\beta_3$ & $\rm-8.2\times10^{-3}ps^3/m$ \\  
             \hline
              $\beta_4$ & $\rm3.1\times10^{-4}ps^4/m$  \\  
              \hline
            \end{tabular}
    \label{tab:s1}
\end{table}

\section{Mapping from the frequency domain to the time domain}\label{A2}
\begin{figure}[!h]
\centering
    \includegraphics[width=0.6\textwidth]{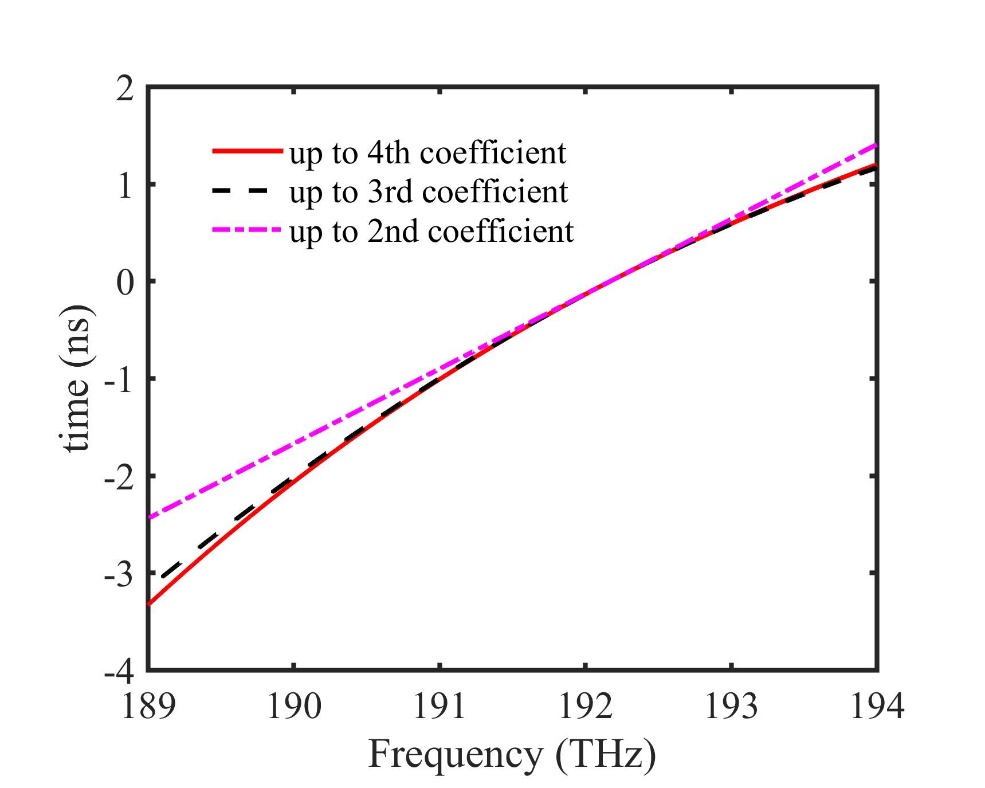}
    \caption{Relation between the optical frequency and the time with 300 m fiber length, taking into account the different orders of dispersion coefficients.}
    \label{fig:s3}
\end{figure}

\noindent The relation between the optical frequency and the time can be obtained by differentiating Eq. (\ref{eq:s2}) with respect to $\omega$. By using the parameters in Table \ref{tab:s1}, we simulate the mapping from the frequency domain to the time domain with 300 m fiber length, and the results with different orders of dispersion coefficients are shown in Fig. \ref{fig:s3}. We find that, if up to $\rm2^{nd}$ order dispersion coefficients are considered in the simulations, the relation between the optical frequency and the time is linear. If higher order dispersion coefficients are involved, the curve is not a straight line.
 
\section{Chirped pulse interferometry with the different fiber lengths}\label{A3}
\noindent As mentioned in the manuscript, the spectrograms will change if the fiber length changes. As shown in Eqs. (2) and (3), the spectrograms corresponding to the reference pulses and the measurement pulses can be expressed as:
\begin{equation} \label{eq:s6}
I_{ref}\left( \omega \right) =E\left( \omega \right) ^2\left[ \alpha ^2+\gamma ^2+2\alpha \gamma cos\!\:\left( \phi _r-\psi \right) \right] 
\end{equation}  
\begin{equation} \label{eq:s7}
I_{meas}\left( \omega \right) =E\left( \omega \right) ^2\left[ \beta ^2+\gamma ^2+2\beta \gamma cos\!\:\left( \phi _r-\tau \omega -\psi \right) \right] 
\end{equation}  
where $\psi(\omega)$ can be calculated as:
\begin{equation} \label{eq:s8}
\psi \left( \omega \right) =\sum\nolimits_{m=1}^{\infty}{\frac{\beta _m}{m!}}\left( \omega -\omega _c \right) ^mz=\beta _1\left( \omega -\omega _c \right) z+\frac{\beta _2}{2}\left( \omega -\omega _c \right) ^2z+\frac{\beta _3}{6}\left( \omega -\omega _c \right) ^3z...
\end{equation}  
We find that, the spectrograms exhibit an oscillation along the horizontal axis (i.e., the angular frequency) due to the cosine terms. We next focus on the cosine term. If the phase of this cosine function is linear, the oscillation of the spectrogram is frequency-stable, which means that the modulation frequency of the spectrogram is constant. This is actually the traditional dispersive interferometry. However, if the phase of the cosine function is not linear, but quadratic, the modulation frequency of the spectrogram is time-varying. In this case, a widest fringe at a specific wavelength will exist, where the phase changes the slowest. This wavelength is also referred to as the balanced wavelength \cite{s1}, and can be used in the stationary phase evaluation \cite{s2}. In our experiments, the pulses are strongly chirped, and there the modulation frequency of the spectrogram is not constant. From Eqs. (\ref{eq:s6}), (\ref{eq:s7}), and (\ref{eq:s8}), it is clear that the modulation frequency will increase when we increase the fiber length $z$. In this case, the width of the widest fringe will decrease correspondingly.

\begin{figure}[!h]
\centering
    \includegraphics[width=1\textwidth]{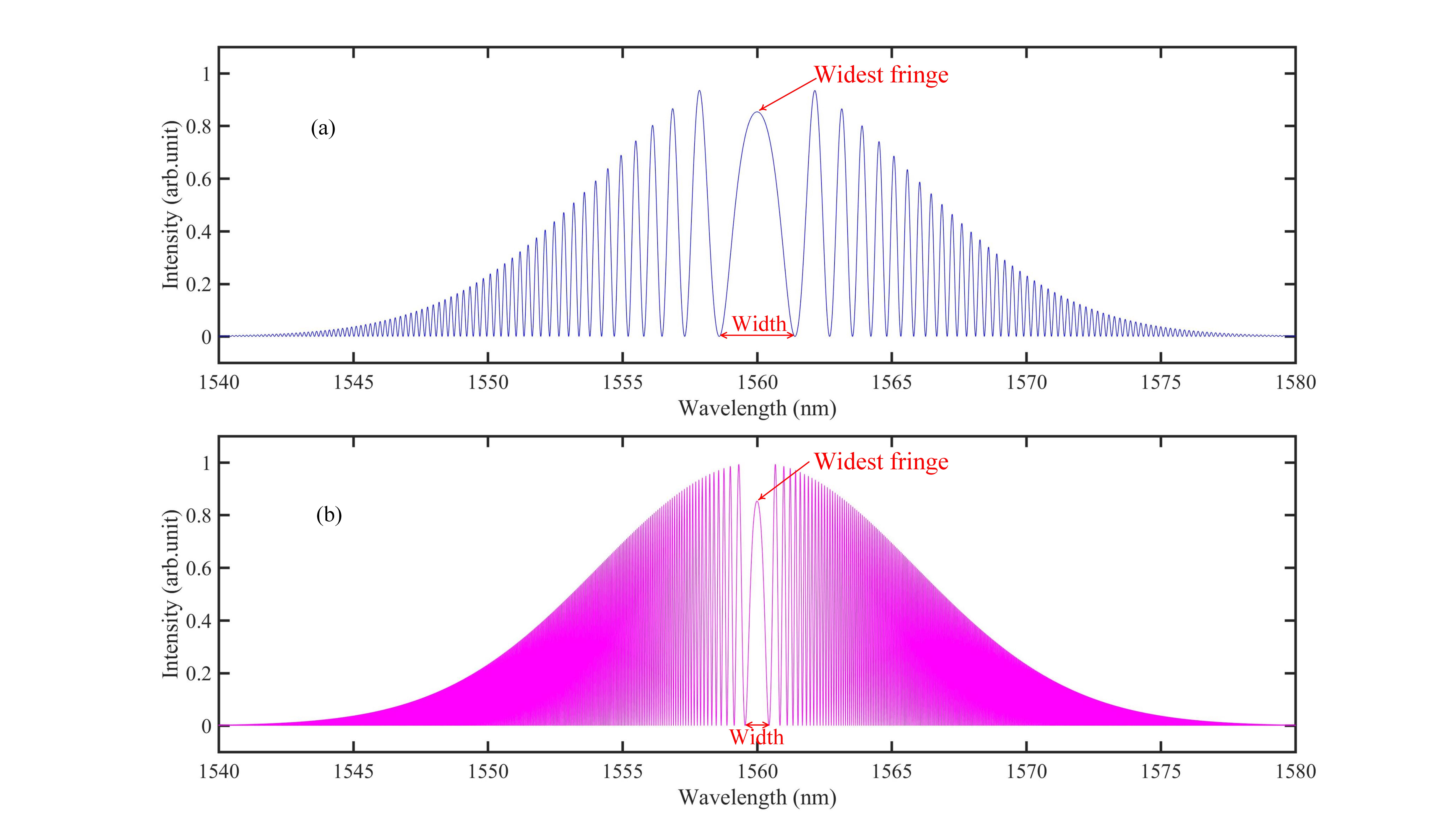}
    \caption{(a): Spectrogram with 50 m fiber length; (b):Spectrogram with 300 m fiber length.}
    \label{fig:s4}
\end{figure}

We carry out the simulations to show the effect of the different fiber lengths, and the results are shown in Fig. \ref{fig:s4}. Fig. \ref{fig:s4}(a) shows the spectrogram when the fiber length is 50 m (i.e., the pulse width after broadening is about 0.4 ns). In contrast, Fig. \ref{fig:s4}(b) indicates the spectrogram when the fiber length is 300 m, and the pulse width is about 4 ns. As discussed before, the modulation frequency of the spectrograms increases when increasing the fiber length, and the width of the widest fringe decreases.
 
In our work, the spectrograms are mapped from the frequency domain to the time domain, so as to greatly improve the measurement speed. Low modulation frequency can relax the requirement of the sampling rate of the oscilloscope. This means that, in the mapping from the frequency domain to the time domain, more fringes can be obtained when the modulation frequency of the spectrogram is lower.
\section{Data process of the chirped pulse interferometry}\label{A4}
\noindent In this section, we describe the data process of the chirped pulse interferometry for distance measurement. When the distances are changed, the fringes are substantially shifted in time. In our experiments, about 8 fringes can be obtained by using the high-speed electronics. We use the fitting phases in the data process, as below.
\begin{figure}[!h]
\centering
    \includegraphics[width=15cm]{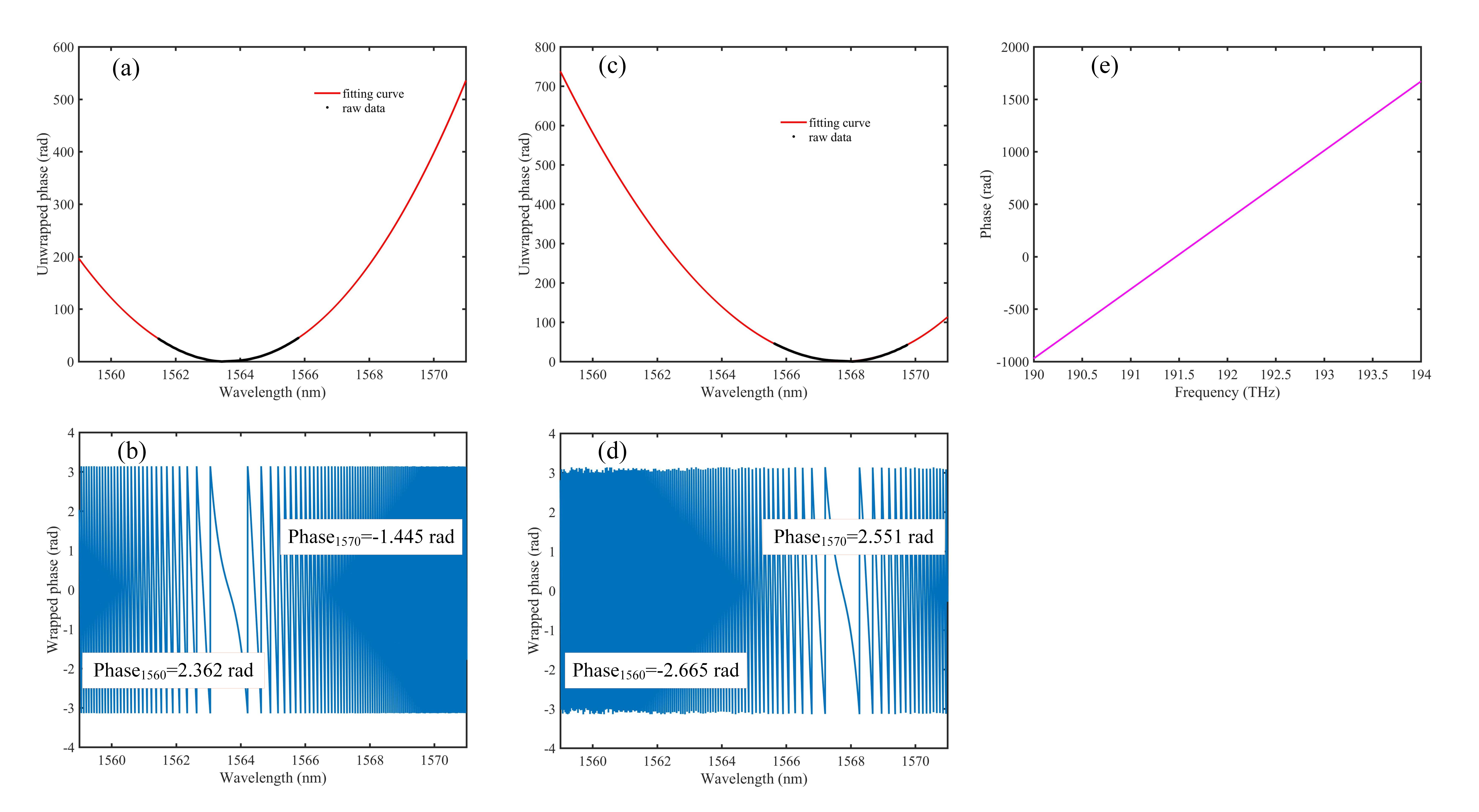}
    \caption{(a): Unwrapped phase of the spectrogram at the initial position; (b): Wrapped phase of the spectrogram at the initial position; The phase changes the slowest at the widest fringe. Here, we can obtain the mode phases at 1560 nm and 1570 nm, respectively, which are 2.362 rad and -1.445 rad. (c): Unwrapped phase of the spectrogram after the distance changes by 15 mm; The position of the inflection point changes accordingly. (d): Wrapped phase of the spectrogram after changing the distance by 15 mm; Similarly, the mode phases at 1560 nm and 1570 nm can be measured, which are -2.665 rad and 2.551 rad, respectively. (e): Phase difference between the unwrapped phases in (a) and (c), which is a straight line with respect to the optical frequency.}
    \label{fig:s9}
\end{figure}

The data process is shown in Fig. \ref{fig:s9}. In Figs. \ref{fig:s9}(a) and \ref{fig:s9}(c), the black solid points represent the phases obtained by the fringes measured by the oscilloscope, i.e., the raw data. However, limited by the sampling rate of the oscilloscope, about 8 fringes can be measured in our experiments, which corresponds to about 4.5 nm band. We then use the curve fitting to derive the phase information out of this band. The red solid lines indicate the fitting curves. We find that, the inflection point is shifted to the “right side” because the distance is changed. The difference between the two unwrapped phases is shown in Fig. \ref{fig:s9}(e), which is a straight line. We can use the slope to calculate the distance which is 15.751 mm. Next, we move to the second step to finely determine the distance. We choose 1560 nm and 1570 nm to generate the synthetic wavelength, which is 244.9 $\upmu$m. The integer part can be calculated as
\begin{equation} \label{eq:s88}
r=round\left( \frac{15.751mm\times 2}{244.9\mu m} \right) =round\left( 64.316\times 2 \right) =128
\end{equation}   
The fractional part can be calculated based on the mode phases shown in Fig. \ref{fig:s9}(d) as
\begin{equation} \label{eq:s888}
e=\frac{2.362+1.445}{2\pi}-\frac{-2.665+2\pi -2.551}{2\pi}=0.436
\end{equation}  
Finally, the distance can be determined as
\begin{equation} \label{eq:s8888}
L=\frac{1}{2}\cdot \left( r+e \right) \cdot \frac{\varLambda}{{n_g}^{\mathrm{'}}}=L=\frac{1}{2}\cdot \left( 128+0.436 \right) \cdot \frac{244.9\mu m}{1.00026737}=15.723mm
\end{equation}

\section{Absolute distance measurement with 300 m dispersion compensation fiber}\label{A5}
\noindent In the manuscript, the length of the dispersion compensation fiber is 90 m. Therefore, the broadening pulse width is only about 1.2 ns, less than 4 ns (i.e., 1/$f_{rep}$). In this case, we should change the repetition frequency to make sure that the spectrograms can always appear. In fact, we realize that the step of the adjustment of the repetition frequency could not be needed if the broadening pulse width can achieve 4 ns.

\begin{figure}[!h]
\centering
    \includegraphics[width=0.9\textwidth]{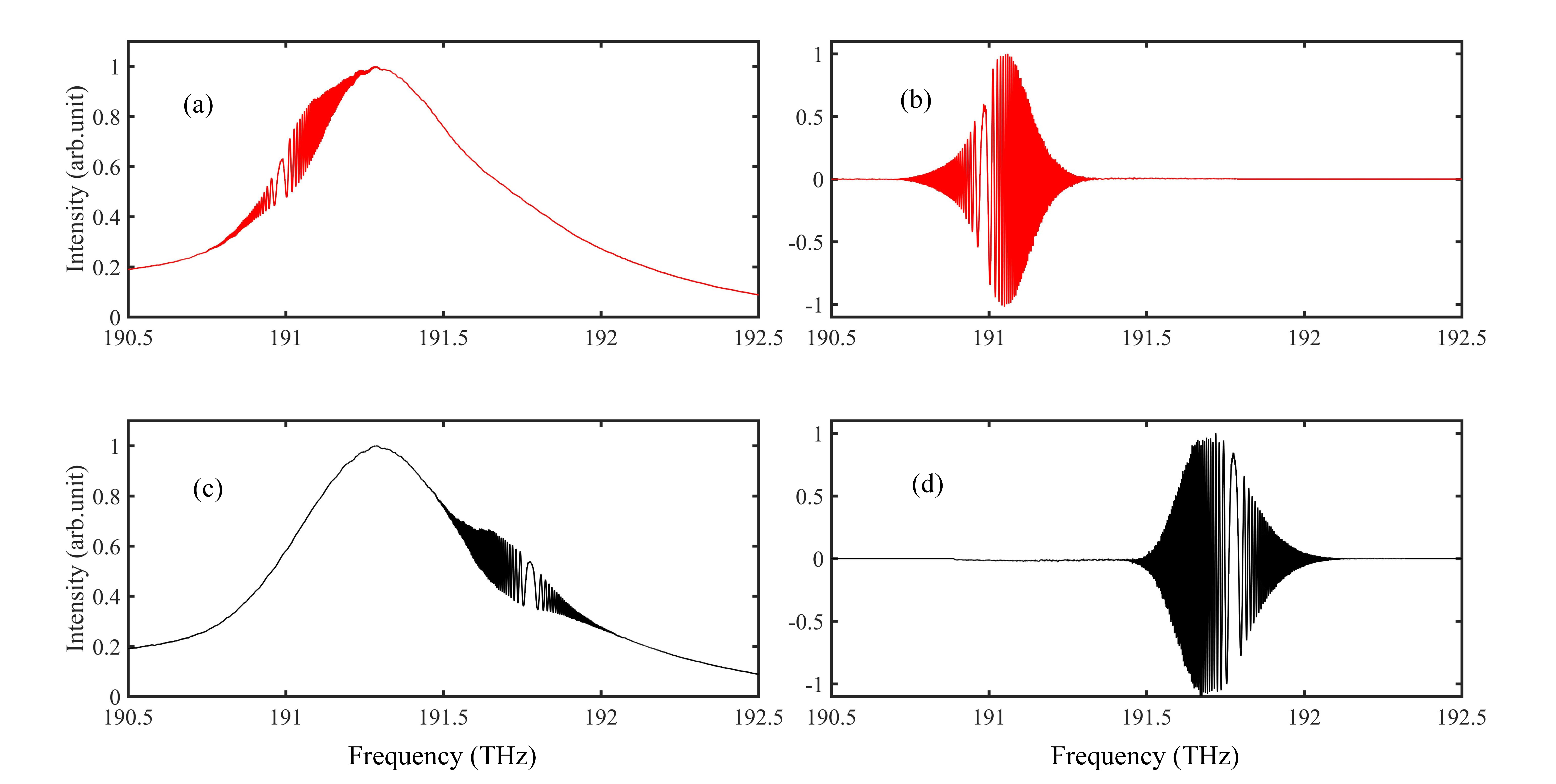}
    \caption{(a): Spectrogram corresponding to the reference mirror. (b): AC part of the spectrogram in (a). (c): Spectrogram corresponding to the target mirror. (d): AC part of the spectrogram in (c).}
    \label{fig:s5}
\end{figure}

Actually, we have carried out this experiments with 300 m DCF. Fig. \ref{fig:s5}(a) indicates the spectrogram corresponding to the reference mirror, and Fig. \ref{fig:s5}(b) shows the AC part. We can clearly find a widest frequency at about 191 THz. Fig. \ref{fig:s5}(c) depicts the spectrogram corresponding to the target mirror, with about 0.25 m distance, and Fig. \ref{fig:s5}(d) is the corresponding AC part. The position of the widest fringe is updated to about 191.8 THz.
 
\begin{figure}[!h]
\centering
    \includegraphics[width=0.8\textwidth]{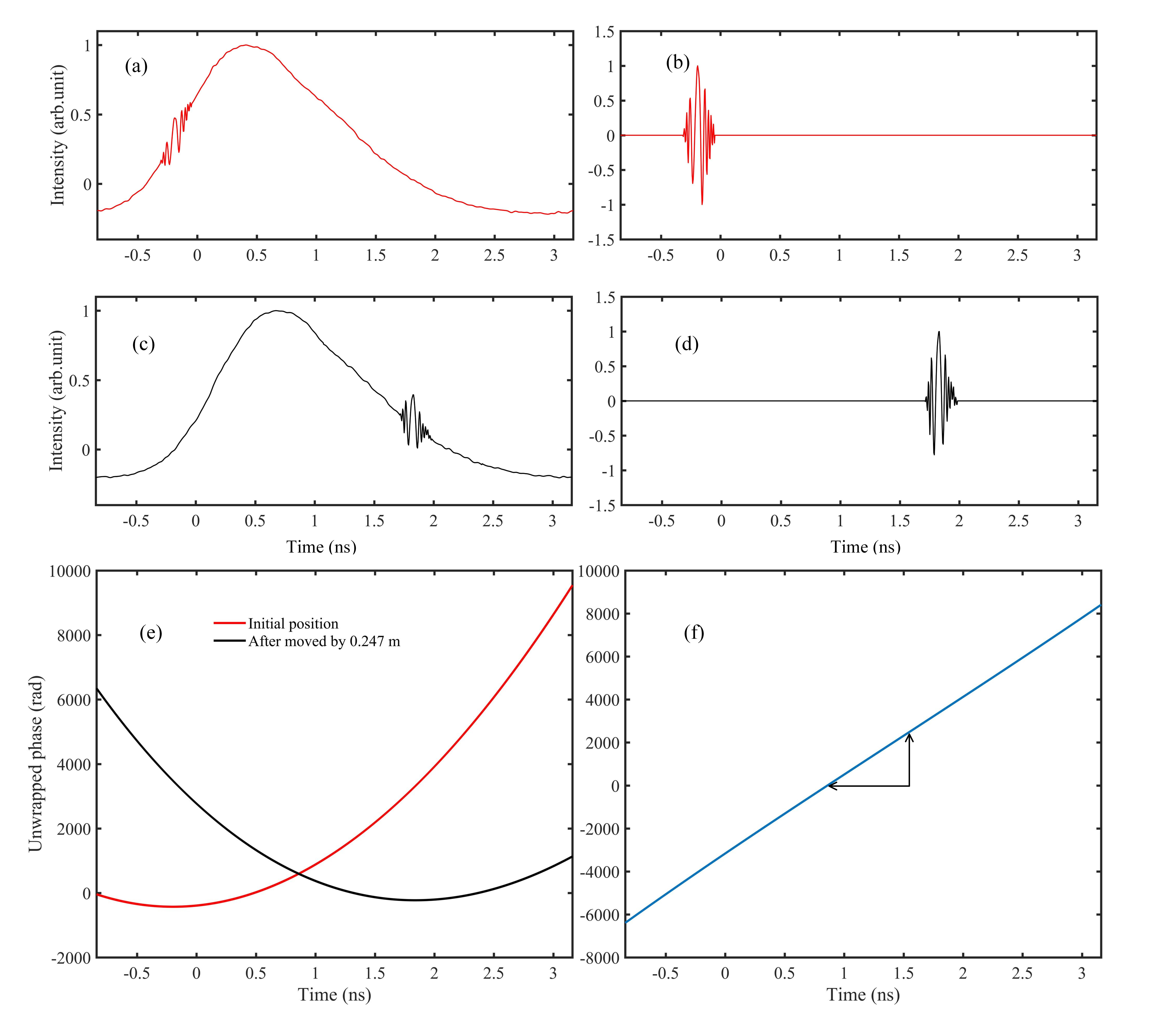}
    \caption{(a): Waveform corresponding to the reference mirror. (b): AC part of the waveform in (a). (c): Waveform corresponding to the target mirror. (d): AC part of the waveform in (c). (e): Unwrapped phase of the spectrograms. The red line indicates the unwrapped phase at the initial position, and the black line represents the phase after moving by 0.247 m. (f): Phase difference between the two phases in (e). We find the phase difference is a straight line, similar with the dispersive interferometry and dual-comb interferometry.}
    \label{fig:s6}
\end{figure}

Fig. \ref{fig:s6} shows the waveforms stored by the oscilloscope, and the data process. After the DC part is removed, the AC part is shown in Figs. \ref{fig:s6}(b) and (d), respectively. We use Hilbert transform to measure the phase of the waveform, and the phases are then unwrapped, as shown in Fig. \ref{fig:s6}(e). The red line indicates the unwrapped phase at the initial position, and the black line represents the phase after the target is moved by 0.247 m. Based on the slope of the phase difference in Fig. \ref{fig:s6}(f), this distance can be determined to 0.24717 m via Eq. (5). The environmental conditions are well controlled, which are 20.9$^{\circ}$C, 1011.1 hPa, and 12.1$\%$ humidity, and the group refractive index is 1.0002683 corrected by the Ciddor formula.
 
\begin{figure}[!h]
\centering
    \includegraphics[width=0.6\textwidth]{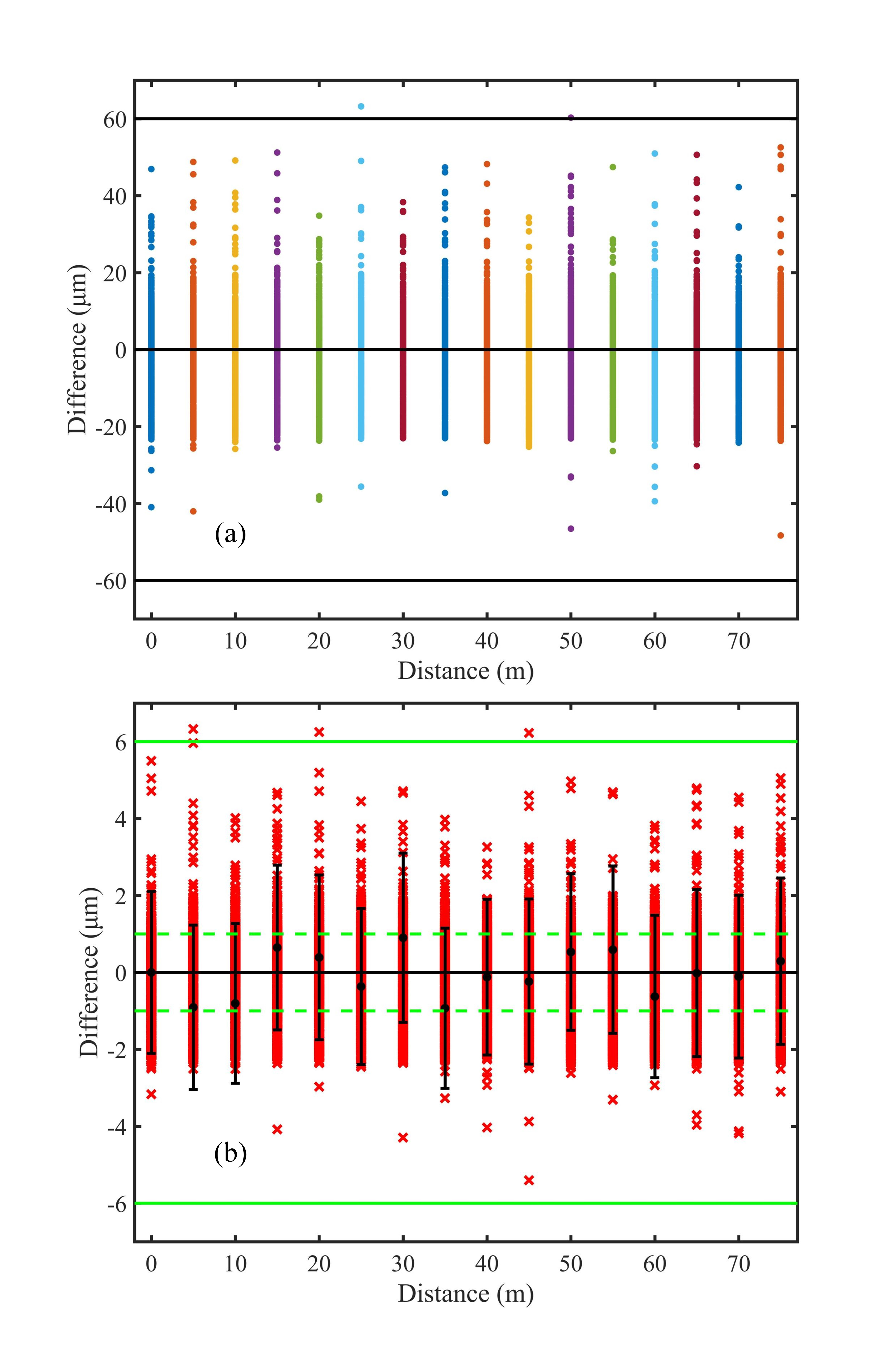}
    \caption{ (a): Results of the coarse measurement; The colorful solid points show the scatters of each individual measurement; (b): Results of the fine measurement. The red x markers show the scatters of 1000 measurements. The black solid points indicate the average value, and the error bar shows twice the standard deviation.}
    \label{fig:s7}
\end{figure}

The results of the distance measurement is shown in Fig. \ref{fig:s7}. The horizontal axis is the reference distances, and the vertical axis is the difference between our measurements and the reference values. Fig. \ref{fig:s7}(a) indicates the results of the coarse measurements. We measure the distances for 1000 times (for just 4 $\upmu$s in fact) at each position, and the colorful solid points are the scatters of each individual measurements. The measuring time for each measurement is 4 ns, corresponding to the repetition frequency of the laser frequency comb. We find that, the measurement uncertainty can be below ±60 $\upmu$m at this step. There is not obviously distance-dependent part in the uncertainty. In the second step, i.e., the fine measurement, the synthetic wavelength should be larger than 2$\times$120 $\upmu$m. We choose the wavelengths of 1565 nm and 1575 nm to generate the synthetic wavelength. Correspondingly, the group refractive index is 1.00026737. The synthetic wavelength is therefore 246.4 $\upmu$m, \textgreater 2$\times$120 $\upmu$m. Based on Eq. (9), the results of the fine measurement are shown in Fig. \ref{fig:s7}(b). The scatters of 1000 measurements are indicated by the red x markers. The black solid points represent the average value of the 1000 measurements, and the error bar shows twice the standard deviation. The green solid lines show the limit of the measurement uncertainty, which is greatly improved to be below ±6 $\upmu$m at 75 m range, corresponding to $8\times10^{-8}$ in relative. We find that, the average values can be well below ±1 $\upmu$m, which is actually $1.3\times10^{-8}$ in relative.
 
\begin{figure}[!h]
\centering
    \includegraphics[width=0.6\textwidth]{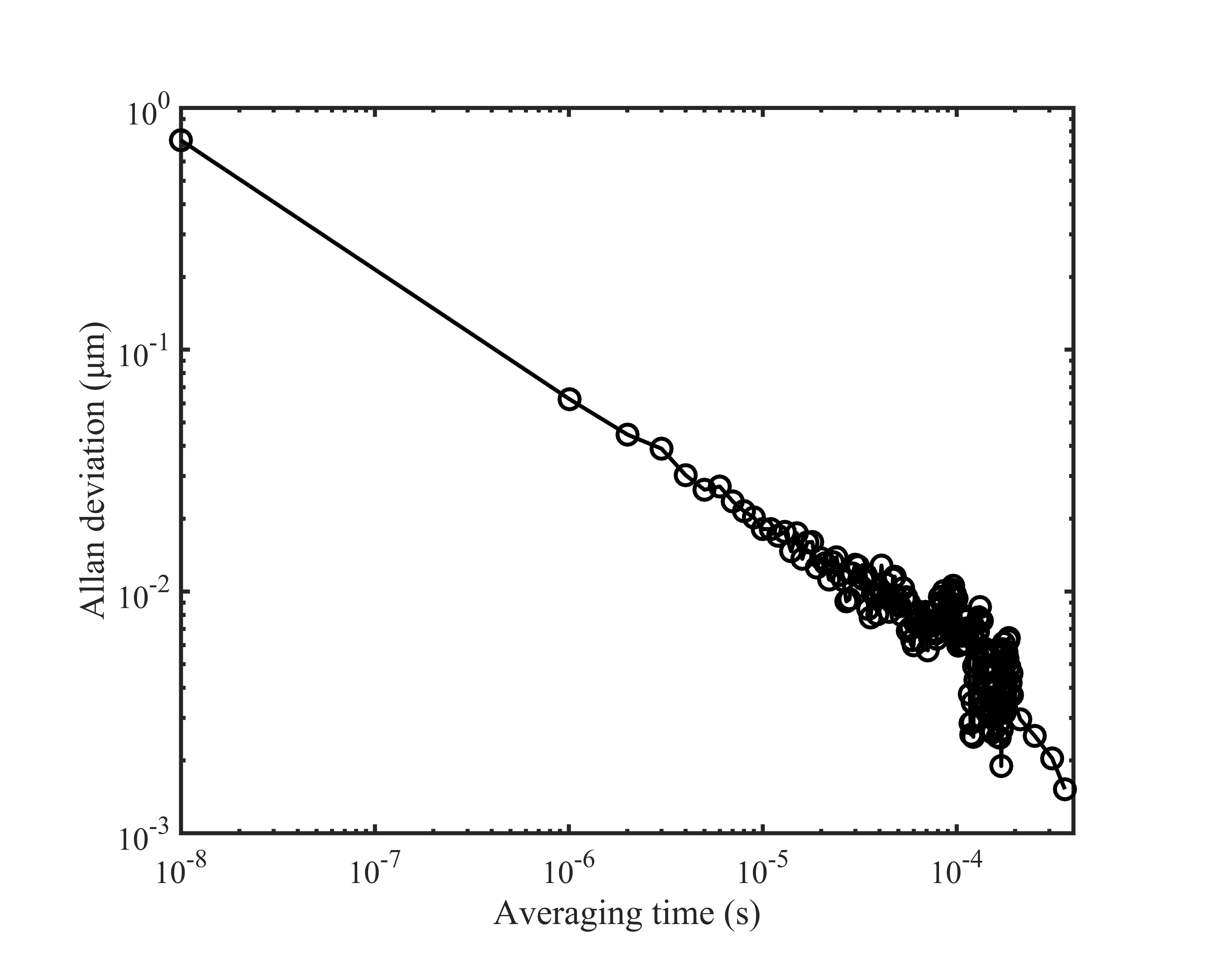}
    \caption{Allan deviation at different averaging time.}
    \label{fig:s8}
\end{figure}

We carry out long-term experiments for 800 $\upmu$s with 200000 measurements, to examine the precision limit. The standard deviation is 1.038 $\upmu$m with averaging time of 4 ns. Fig. \ref{fig:s8} shows the Allan deviation at 50 m distance. The Allan deviation is 0.7 $\upmu$m at averaging time of 10 ns, 18 nm at 10 $\upmu$s, and can achieve 1.5 nm at 360 $\upmu$s averaging time. These results show that nanometer-level precision can be realized.

\end{appendices}


\bibliographystyle{IEEEtran}
\bibliography{IEEEabrv,sample}

\newpage

\end{document}